\documentclass{aa}
\usepackage{graphicx}
\usepackage{txfonts}
\usepackage{epsfig,graphics,graphicx,bm,amssymb}
\usepackage[normalem]{ulem}

\usepackage[dvipsnames]{xcolor}

\usepackage[]{natbib}
\usepackage[hidelinks]{hyperref}

\begin{document}

\title{Modelling resonances and orbital chaos in disk galaxies. Application to a Milky Way spiral model}

\author{T.\,A. Michtchenko\thanks{e-mail: tatiana@astro.iag.usp.br}
          \and \, R.\,S.\,S. Vieira\thanks{e-mail: rss.vieira@usp.br}
          \and \, D.\,A. Barros\thanks{e-mail: douglas.barros@iag.usp.br}
          \and \,J.\,R.\,D. L\'epine\thanks{e-mail: jacques@astro.iag.usp.br}
}

\institute{Instituto de Astronomia, Geof\'isica e Ci\^encias Atmosf\'ericas, USP, Rua do Mat\~ao 1226, 05508-090 S\~ao
Paulo, Brazil}

\date{}

\abstract
{Resonances in the stellar orbital motion under perturbations from the spiral arm structure can play an important role in the evolution of the disks of spiral galaxies. The epicyclic approximation allows the determination of the corresponding resonant radii on the equatorial plane (in the context of nearly circular orbits), but is not suitable in general.
}
{We expand the study of resonant orbits by analysing stellar motions perturbed by spiral arms with Gaussian-shaped groove profiles without any restriction on the  stellar orbital configurations, and we expand the concept of Lindblad (epicyclic) resonances for orbits with large radial excursions.
}
{We define a representative plane of initial conditions, which covers the whole phase space of the system. Dynamical maps on representative planes of initial conditions are constructed numerically in order to characterize the phase-space structure and identify the precise location of the co-rotation and Lindblad resonances. The study is complemented by the construction of dynamical power spectra, which provide the identification of fundamental oscillatory patterns in the stellar motion.}
{Our approach allows a precise description of the resonance chains in the whole phase space, giving a broader view of the dynamics of the system when compared to the classical epicyclic approach. We generalize the concept of Lindblad resonances and extend it to cases of resonant orbits with large radial excursions, even for objects in retrograde motion. The analysis of the solar neighbourhood shows that, depending on the current azimuthal phase of the Sun with respect to the spiral arms, a star with solar kinematic parameters (SSP) may evolve in dynamically distinct regions, either inside the stable co-rotation resonance or in a chaotic zone.}
{Our approach contributes to quantifying the domains of resonant orbits and the degree of chaos in the whole Galactic phase-space structure. It may serve as a starting point to apply these techniques to the investigation of clumps in the distribution of stars in the Galaxy, such as kinematic moving groups.}


\titlerunning{Stellar dynamics in spiral galaxies}

\maketitle

\section{Introduction}\label{intro}

A general consensus on the spiral arm structure of the Galaxy has not yet been reached. Questions like the physical nature of the arms,  their exact location,  the existence and position of resonances,  their role in the evolution of the galactic disk, and the connection of the spiral arms with the bar have originated papers with quite diverse views.  One  of the main  questions which divides the astronomical community  is  the lifetime of the spiral structure. Many authors consider that the spiral arms of a grand-design spiral are long-lived, with lifetimes of a few billion years, and rotate like a rigid body (e.g. \citealp{bertin1996spiral} and references therein). The first model of spiral arms  proposed by \citet{linShu1964ApJ}, in which the stars were treated as a fluid, adopted this view. However, the more recent ``long-lived arms'' models are based on a very different understanding  of the nature of the arms. They focus on stellar orbits, and have adopted a concept introduced by \citet{kalnajs1973PASAu}, who considers that the arms are places where neighbouring galactic orbits of stars become close one to the other, resulting in regions of high stellar density or elongated spiral-shaped potential wells.

Considerable effort has been devoted to the self-consistency of models    \citep[e.g. ][among others]{contopoulosGrosbol1986AA, amaralLepine1997MNRAS, pichardoEtal2003ApJ, martosEtal2004MNRAS, junqueiraEtal2013AA} that we briefly explain below. Series of closed stellar orbits are calculated in a potential which is the sum of the axisymmetric potential of the disk plus an imposed spiral-shaped perturbation. These orbits naturally tend to present concentrations (high proximity) in some regions and, therefore, to produce new potential perturbations. If these newly created perturbations have a shape similar to the one that was originally imposed then the organization of the orbits will persist for long periods, so that the spiral structure is long-lived or self-consistent. The fact that the above studies show that such self-consistent solutions do exist justifies that it is an acceptable approximation to impose a spiral-shaped perturbation in order to perform studies of stellar orbits without worrying  about the origin of this constant perturbation any more.

Most of the models aiming to describe stellar orbits adopt a perturbation potential given by a cosine law in the azimuthal direction, as explained in more detail in this paper. This comes from a long tradition and for simplicity. However, this potential is too smooth, with broad maxima and minima, while the density of orbits that are obtained vary sharply.  \citet{junqueiraEtal2013AA} proposed a new expression in which the arms are Gaussian-shaped wells, or grooves, in the azimuthal direction. This model results in a much better self-consistency since the width of the grooves derived from the stellar density is equal to that of the imposed perturbation.  In the present work we adopt the potential of \citet{junqueiraEtal2013AA}, re-scaled to new parameters $R_0$ and $V_0$.

Another line of research  which is current in the literature, namely  N-body numerical simulations \citep[e.g. ][among others]{sellwoodCarlberg1984ApJ, sellwoodKahn1991MNRAS}, is quite distinct and competes with the one that we adopt in this paper. In general, these studies find that spiral arms are transient, appearing and disappearing in a recurrent way. In these studies, the positions of resonances are not well defined as they can move with time, and different parts of the structure can have, for instance, different co-rotation radii.

We believe, however, that there is evidence that the co-rotation resonance usually stays at a same radius for a few billion years. This is suggested, for instance, by the step in metallicity at co-rotation in our Galaxy \citep[see e.g. ][their Figure 4]{lepineEtal2011MNRAS}, and the breaks in the gradients of metallicity at co-rotation, in external galaxies \citep{scaranoLepine2013MNRAS}.

Regarding the dynamical modelling of stellar orbits in spiral galaxies, most of the studies analyse the dynamics of stars on the equatorial plane of galaxy models when subjected to small azimuthal perturbations (such as central bars and spiral arms). Orbits in barred galaxies are reviewed in \citet{contopoulosGrosbol1989AARev} and \citet{contopoulosEtal1996LNP}. Recently, numerical studies of chaos in long-lived spiral and barred galaxies have been performed in \citet{pichardoEtal2003ApJ, pichardoMartosMoreno2004ApJ, chakrabartySideris2008AA, contopoulos2009CeMDA, patsis2012IJBC}, and \citet{morenoPichardoSchuster2015MNRAS}, among others. Prospects of observing chaos in disk galaxies are summarized in \cite{grosbol2002SSRv, grosbol2003LNP, grosbol2009}. Also, many recent studies have focused on the phase-space structure of the solar neighbourhood \citep{dehnen2000AJ, quillen2003AJ, chakrabarty2004MNRAS,  chakrabarty2007AA, pichardoMartosMoreno2004ApJ,  chakrabartySideris2008AA, antojaEtal2008AA, antojaEtal2009ApJL, antojaEtal2011MNRAS, morenoPichardoSchuster2015MNRAS}.

The present work, which follows the approach described above, is a theoretical and numerical study of dynamical consequences of a long-lived spiral structure, such as the presence of resonances and regions of chaotic orbits, in a galaxy described by the spiral potential perturbation proposed by \citet{junqueiraEtal2013AA}. We obtain dynamical maps and dynamical power spectra for the orbits restricted to the equatorial plane, which are tools widely used in celestial mechanics \citep[e.g. ][]{michtchenkoEtal2002Icar, ferrazmeloMichtchenkoEtal2005LNP}, and compare the results  with analytical results for the axisymmetric potential. In particular, we show that our approach describes the positions of the resonance chains in the whole phase space with great precision, expanding the concept of Lindblad (epicyclic) resonances for orbits with large radial excursions, and for the case of retrograde orbits.

The model that we generate is useful in order to understand the observed features of our Galaxy, at least in the range of radii where the adopted perturbation seems to be similar to the observed spiral arms. This range should exclude the inner 3 kpc, since the adopted perturbation does not consider the presence of a bar, and possibly also the outer regions of the Galaxy, since we do not know up to what distance the potential perturbation that we adopt is a good description of the real one. We emphasize that the present study deals only with resonance effects induced by the spiral perturbation. Therefore, any possible resonant structures influenced by the central bar will not be considered in the present analysis.

The outline of the paper is as follows.  In Sect.\,2, using observational data, we develop a rotation-curve model to obtain the axisymmetric gravitational potential. In Sect.\,3 we include the perturbing potential of spiral galaxies  in the Hamiltonian model, and in Sect.\,4 we briefly introduce the analytical and numerical techniques used in this study. The stationary solutions of the Hamiltonian are obtained and analysed in Sect.\,5, where we also introduce the concept of spiral branches. Section\,6 is devoted to the detailed analysis of the topology of the Hamiltonian on the representative plane. Sect.\,7 presents dynamical maps of the phase space of the system under study, together with dynamical power spectra, which allow us to identify the main Lindblad resonances. The comparison of our results with those obtained via the epicyclic approximation is also done in this section. In Sect.\,8, we briefly discuss the influence of the co-rotation dip in the rotation curve on the stellar dynamics. Finally, the discussions and conclusions are presented in Sect.\,9, while Appendices A and B describe two additional points, namely the comparison between the Gaussian and cosine profiles of the spiral arms and the detailed description of the tools and methods employed in this study.


\section{Rotation curve and axisymmetric potential}\label{sec:rotcurve}

In this paper, we consider a realistic rotation-curve model of the Milky Way  based on published observational data. For the Galactocentric distance of the Sun, we adopt \mbox{$R_{0}=8.0$\,kpc}, which is based on the statistical analysis performed by \citet{malkin2013IAUS} on several $R_0$ measurements published in the literature. The circular velocity at $R_0$, which is the velocity $V_0$ of the local standard of rest (LSR), is chosen to satisfy the relation \mbox{$V_{0}=R_{0}\Omega_{\odot}-v_{\odot}$}, where \mbox{$\Omega_{\odot}=30.24$\,km\,s$^{-1}$\,kpc$^{-1}$} is the angular rotation velocity of the Sun (\citealt{reidBrunthaler2004ApJ}), and \mbox{$v_{\odot}=12.24$\,km\,s$^{-1}$} is the peculiar velocity of the Sun in the direction of Galactic rotation (\citealt*{schonrichBinneyDehnen2010MNRAS}). These values result in \mbox{$V_{0}=230$\,km\,s$^{-1}$}.

For the rotation curve, we use the tangent-point data in H\,{\scriptsize I} from \citet{burtonGordon1978AA} and  \citet*{fichBlitzStark1989ApJ}, and the CO-line tangent-point data from \citet{clemens1985ApJ}. We also use data of maser sources associated with high-mass star-forming regions, obtained from Table 1 in \citet{reidEtal2014ApJ}. From  these compiled data, the rotation velocities and Galactic radii were calculated using the Galactic constants \mbox{($R_0$, $V_0$)} adopted in this work.

\begin{figure}
\begin{center}
\epsfig{figure=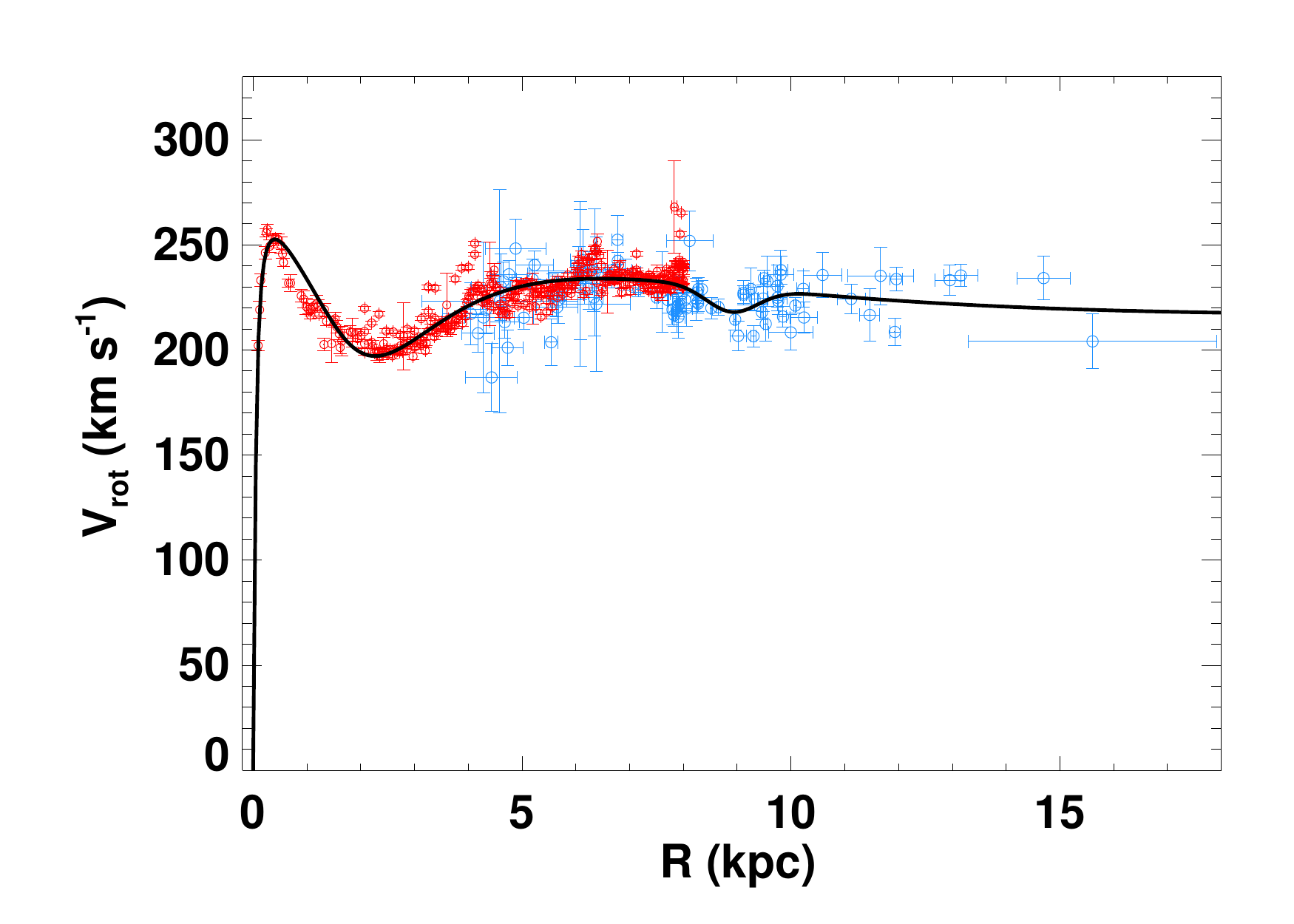,width=0.99\columnwidth ,angle=0}
\caption{Rotation curve of the Galaxy. Red points indicate  H\,{\scriptsize I} and CO tangent-point data from \citet{burtonGordon1978AA}, \citet{clemens1985ApJ}, and \citet{fichBlitzStark1989ApJ}; blue points indicate  masers from high-mass star-forming regions from \citet{reidEtal2014ApJ}. The black curve indicates the fitted rotation curve expressed by Eq.\,(\ref{eq:Vrot}).
}
\label{fig:Vrot}
\end{center}
\end{figure}

Figure \ref{fig:Vrot} shows the rotation curve of the Galaxy built up with the observational data described above: red points represent the H\,{\scriptsize I} and CO tangent-point data, while blue points correspond to the maser sources data. In order to obtain a realistic model for the rotation curve, we fit the observational data by a convenient expression given by the sum of three exponentials in the form

\begin{eqnarray}
\label{eq:Vrot}
 V_{\rm rot}(R) &=& 302\,\exp\left(-\frac{R}{4.53}-\frac{0.036}{R}\right)  \nonumber\\
 & & + 223\,\exp\left[-\frac{R}{1959}-\left(\frac{3.64}{R}\right)^2\right] \\
               & &   -11.86\,\exp\left[-\frac{1}{2}\left(\frac{R-8.9}{0.48}\right)^2\right],  \nonumber
\end{eqnarray}
\noindent
with the factors that multiply the exponentials given in units of kilometers per second, and the factors in the arguments of the exponentials given in kiloparsecs.
The numerical values of the model rotation curve in Eq.\,(\ref{eq:Vrot}) were obtained after  fitting  the observational data: We minimized the sum of the squares of the residuals between $V_{\rm rot}$ and the measured rotation velocities, weighted by their respective uncertainties.

The smooth black curve in Fig.\,\ref{fig:Vrot} represents the rotation velocity given by Eq.\,(\ref{eq:Vrot}). It is worth noting a dip in the rotation curve  centred at 8.9 kpc with a super-Keplerian fall-off. This velocity dip (represented by the third term in Eq.\,(\ref{eq:Vrot}))  is related to a local minimum followed by a local maximum in the disk's surface density distribution, as explained in Sect.\,\ref{sec:corot_dip} (see also \citealt*{barrosLepineJunqueira2013MNRAS}). However, the surface density in this range always remains  positive, which would not be the case if  the spherical version of Eq.\,(\ref{eq:Vrot}) is inadvertently used.

The radial gradient of the axisymmetric potential $\Phi_{0}(R)$ is related to the rotation velocity $V_{\rm rot}(R)$ through
\begin{equation}\label{eq3}
F=-\frac{\partial\Phi_0}{\partial R} = - \frac{V^2_{\rm rot}}{R},
\end{equation}
where $V_{\rm rot}$ is given by the rotation curve from Eq.\,(\ref{eq:Vrot}). To obtain the axisymmetric potential $\Phi_0(R)$,  we use  the trapezium rule with adaptive step to solve numerically the integral of Eq.~(\ref{eq3}). The constant of integration is formally obtained from the limit condition \mbox{$ \displaystyle\lim_{R \to +\infty} \Phi_0 = 0$}. In practice, $\infty$ can be replaced by a large value of $R$; for instance, 1000\,kpc is found to be a good choice. We note that the maximum range of applicability of our model is constrained by a radius of \mbox{$\sim 30$}\,kpc,  as we show in Sect.\,\ref{sec:maps}.

\section{Model}\label{model}

\citet{junqueiraEtal2013AA} proposed a new description of the perturbed gravitational potential of spiral galaxies where the spiral arms have Gaussian-shaped groove profiles. In that approach, the surface density of a zero-thickness disk is represented analytically as the sum of an axisymmetric (unperturbed) surface density $\Sigma_0(R)$ plus a small perturbation $\Sigma_1(R,\varphi)$, which describes the spiral pattern in a rotating frame with angular speed $\Omega_p$. The azimuthal coordinate in the rotating frame is \mbox{$\varphi=\theta-\Omega_p\,t$}, where $\theta$ is the angular coordinate with respect to  the inertial frame.

The Hamiltonian which describes the stellar dynamics on the equatorial plane under perturbations of the spiral galaxy potential is written as
\begin{equation}\label{eq1}
{\mathcal H}(R,\varphi,p_r,J_\varphi)= {\mathcal H_{0}}(R,p_r,J_\varphi) + \Phi_{1}(R,\varphi),
\end{equation}
with ${\mathcal H_{0}}$ and $\Phi_{1}$ being the unperturbed and perturbation components, respectively. The momenta $p_r$ and \mbox{$J_\varphi=p_\theta$} are the linear and angular momenta per unit mass, respectively. It is worth noting that $J_\varphi$ is measured with  respect to the inertial frame, but it is also a conjugate momentum to the canonical coordinate $\varphi$ of the rotating frame:
\mbox{$J_\varphi=R^2\dot\theta=R^2(\dot\varphi+\Omega_p)=p_\varphi$}.

The one-degree-of-freedom unperturbed Hamiltonian is given by Jacobi's integral
\begin{equation}\label{eq:H0}
{\mathcal H_{0}}(R,p_r,J_\varphi)= \frac{1}{2}\left[ p_r^2 + \frac{J_\varphi^2}{R^2}\right] - \Omega_p J_\varphi + \Phi_0(R),
\end{equation}
where $\Phi_0(R)$ is the galactic axisymmetric potential obtained in Sect.\,\ref{sec:rotcurve}. In the expression above, the first term defines the kinetic energy $T$ of a star and the second is a gyroscopic term.

\begin{table}
\caption{Adopted spiral arms parameters.}
\label{tab:1}       
\begin{tabular}{lclc}
\hline 
\hline
 & & &  \\
Parameter & Symbol & Value & Unit  \\
\hline
Number of arms        & m                   & 2         & -   \\
Pitch angle           & i                   & -14$^\circ$& -   \\
Arm width            & $\sigma$            & 5.0       & kpc \\
Scale length          & $\varepsilon_s^{-1}$& 2.7       & kpc  \\
Spiral pattern speed  & $\Omega_p$          & 26        & km\,s$^{-1}$\,kpc$^{-1}$\\
Perturbation amplitude& $\zeta_0$           & 630       & km$^2$\,s$^{-2}$\,kpc$^{-1}$\\
Reference radius              & $R_i$               & 8                    & kpc\\
\hline
\end{tabular}
\end{table}

The perturbation component of the Hamiltonian (\ref{eq1}) is defined in \citet{junqueiraEtal2013AA} as
\begin{equation}\label{eq:H1gaussian}
 \Phi_1(R,\varphi) = -\zeta_0\,R\,e^{-\frac{R^2}{\sigma^2}[1-\cos(m\varphi-f_m(R))]-\varepsilon_s R},
\end{equation}
where $\zeta_0$ is the perturbation amplitude, $\varepsilon_s^{-1}$ is the scale length of the spiral, $\sigma$ is the width of the Gaussian profile  in the galactocentric azimuthal direction, and $f_m(R)$ is the shape function given by
\begin{equation}\label{eq6}
f_m(R) = \frac{m}{\tan(i)}\ln{(R/R_i)}+ \gamma,
\end{equation}
where $m$ is the number of arms fixed at 2 in our model, $i$ is the pitch angle, $R_i$ is an arbitrary radius chosen to adjust the phase of the spirals (here we adopt \mbox{$R_{i}=R_{0}$)}, and $\gamma$ is a phase angle, which does not influence the dynamics and can be initially fixed at 0.
It is worth noting that the rotation curve of the Galaxy was fitted without analysing the mass components that explain it, and a kind of axisymmetric average of the spiral structure is one of the mass components that are naturally included in the fit. When we add the perturbation potential to the axisymmetric potential (Eq.\,(\ref{eq:H0})), the effect of the arms is counted a second time. The function of Eq.\,(\ref{eq:H1gaussian}) corresponds to a positive density at all points, so that we are increasing the total mass of the disk. We estimate that the mass of the spiral arms corresponding to the potential of Eq.\,(\ref{eq:H1gaussian}) is of the order of 5\% of the mass of the disk. Following \citet{junqueiraEtal2013AA}, we consider that not removing the contribution of the spiral arms from the unperturbed Hamiltonian before going to Eq.\,(\ref{eq:H0}) is a valid approximation.

\begin{figure}
\begin{center}
\epsfig{figure=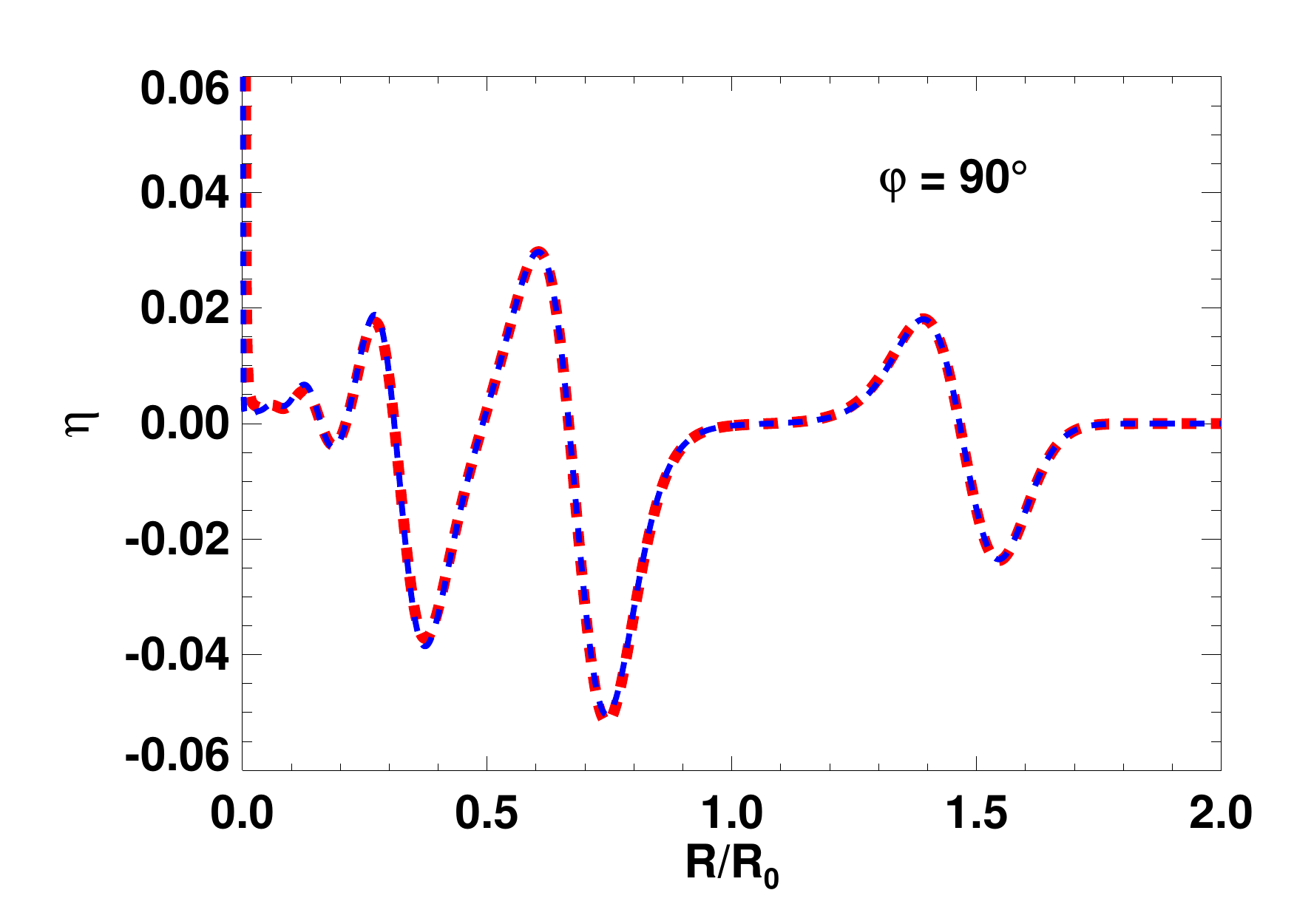,width=0.99\columnwidth ,angle=0}
\caption{Variation of the ratio \mbox{$\eta=(\partial \Phi_1/\partial R)/(\partial \Phi_0/\partial R)$} as a function of normalized Galactic radius $R/R_0$ and along the azimuth \mbox{$\varphi=90^{\circ}$}, for the potential model of \citet{junqueiraEtal2013AA} (red dashed curve) and for the potential model with updated parameters adopted in the present work (blue dashed curve). $\Phi_0$ and $\Phi_1$ are the unperturbed and perturbed parts of the Galactic potential, respectively.}
\label{fig:eta}
\end{center}
\end{figure}

The parameters in Eq.\,(\ref{eq:H1gaussian}) are given in Table\,\ref{tab:1}. Their choice was based on the adopted rotation curve with Galactic constants         \mbox{$R_0=8$\,kpc} and \mbox{$V_0=230$\,km\,s$^{-1}$}.  Since these constants are different from those used by \citet{junqueiraEtal2013AA} (\mbox{$R_0=7.5$\,kpc}, \mbox{$V_0=210$\,km\,s$^{-1}$} in that case), the parameters of the axisymmetric potential $\Phi_0(R)$ are also different from the previous ones. To keep  the self-consistency of the spiral perturbation model of \citet{junqueiraEtal2013AA} unaltered, the parameters of the perturbed Hamiltonian in Eq.\,(\ref{eq:H1gaussian}) also needed to be updated. To do so, we considered that the ratio between the radial forces due to the spiral perturbation and due to the axisymmetric potential is a quantity that has to be preserved after the re-scaling of $\Phi_0(R)$ with the new pair of constants ($R_0$, $V_0$). In other words, we searched for the correspondence between the ratio \mbox{$\eta=(\partial \Phi_1/\partial R)/(\partial \Phi_0/\partial R)$} calculated with the parameters of the Junqueira et al. model and that calculated with the parameters adopted in the present work. Here, $\Phi_1$ is the potential due to the perturbation, which is given by the expression in Eq.\,(\ref{eq:H1gaussian}). Figure \ref{fig:eta} shows the comparison between the ratio $\eta$ calculated with the potential model of \citet{junqueiraEtal2013AA} (red dashed curve) and the ratio obtained with our updated models (blue dashed curve), in the direction of the azimuth $\varphi=90^{\circ}$ for illustration. To make clear the correspondence between the curves, Fig.\,\ref{fig:eta} shows the ratio $\eta$ as a function of Galactocentric radius normalized with respect to $R_0$. This correspondence was obtained by setting new values for the amplitude $\zeta_0$, the scale length $\varepsilon_{s}^{-1}$, and the width $\sigma$ of the spiral potential model, which are presented in Table\,\ref{tab:1}.

The angular speed of the spiral pattern was directly measured by \citet{diasLepine2005ApJ}, using the birthplaces of samples of observed open clusters. The authors found \mbox{$\Omega_p=25\pm 1$\,km\,s$^{-1}$\,kpc$^{-1}$} (see also the review by \citealt{gerhard2011MSAIS} on the values of $\Omega_p$ estimated in the literature). In this work, we adopt \mbox{$\Omega_p=26$\,km\,s$^{-1}$\,kpc$^{-1}$}. For the rotation curve of Eq.\,(\ref{eq:Vrot}), it results in the co-rotation radius \mbox{$R_{CR}=8.54$\,kpc} estimated from the condition \mbox{$V_{\rm rot}(R_{CR})=R_{CR}\,\Omega_p$}. This value for the co-rotation radius is in agreement with the ratio \mbox{$R_{CR}/R_0=1.06$} as determined in \citet{diasLepine2005ApJ}.

\paragraph{Cosine spiral pattern.}\label{sec:cos-spiral}

We also work in this paper with the widely used cosine perturbation given by
        \begin{equation}\label{eq:PhiCosine}
          \Phi_{1,\rm cos}(R,\varphi) = -\zeta_0 R e^{-\varepsilon_S R}\cos{[m\varphi-f_m(R)]}.
        \end{equation}
Our goal is to compare the dynamics under two different perturbations, Gaussian (\ref{eq:H1gaussian}) and cosine, which will be done in Appendix \ref{app1}. We advance that the two potentials do not present significant qualitative differences in the dynamics, except near the main Lindblad resonances and for large radii.


\section{Tools and methods}\label{sec:tools}

We briefly present in this section the tools employed in the orbital analysis throughout the paper. First we present the analytical tools, and then we describe the numerical techniques for detecting resonances and chaos.

The analytical study of motion is done in the context of the unperturbed Hamiltonian, described by the term $\mathcal {H}_{0}(R,p_r,J_\varphi)$ in Eq.\,(\ref{eq1}). In particular, the two independent frequencies are estimated; they are given as
\begin{equation}\label{eq:fphifR}
 f_\varphi = \frac{1}{T_\varphi} = \frac{|\Omega_\varphi|}{2\pi}, \qquad
 f_R = \frac{1}{T_R} = \frac{\Omega_R}{2\pi},
\end{equation}
where $T_\varphi$ and $T_R$ are the azimuthal and radial periods in the rotating frame, respectively,  and $\Omega_\varphi$ and $\Omega_R$ are the corresponding angular frequencies. We then have \mbox{$\Omega_\varphi=\Omega_\theta-\Omega_p$}, where \mbox{$\Omega_\theta=2\pi/T_{\theta}$} is the orbital frequency in the inertial frame. Resonances are then given by
\begin{equation}\label{eq:resonances}
f_\varphi=\frac{j}{n}\,f_R,
\end{equation}
with $j$ and $n$ coprime integers.

The orbital ($\theta$-) and radial ($R$-) periods of a regular bounded orbit are estimated using the axially symmetric potential $\Phi_0$
\citep[see also Appendix\,\ref{sec:frequenciesUnperturbed}]{binneytremaineGD}.
In a frame rotating with angular velocity $\Omega_p$, the azimuthal coordinate $\varphi$ has a variation $\Delta\varphi$ along one radial period of the orbit. The positions of resonances calculated from the unperturbed problem then occur for
\begin{equation}\label{eq:Deltavarphi}
 \frac{\Delta\varphi}{2\pi}=\frac{j}{n}
\end{equation}
(see Appendix\,\ref{sec:frequenciesUnperturbed} for a more detailed discussion). We note that this condition is valid regardless of the approximation of a nearly circular orbit; our calculations are done for general orbits with large radial span.

Regarding numerical techniques, we utilize the Spectral Analysis Method
(SAM; \citealp{michtchenkoEtal2002Icar, ferrazmeloMichtchenkoEtal2005LNP}) in order to analyse the full Hamiltonian~(\ref{eq1}). First, we integrate numerically the equations of motion of stars in order to obtain the corresponding orbits. Each orbit is then Fourier transformed and the number of frequency peaks above a given threshold is calculated (here we adopt 5\% of the amplitude of the highest peak). This number is referred to as \emph{spectral number N}. The spectral number quantifies the chaoticity of each orbit. Regular orbits are conditionally periodic and therefore have a small number of significant frequency peaks (small $N$). On the other hand, chaotic orbits are not confined to an invariant torus. Their frequency spectrum is not discrete and is described by large values of $N$ \citep{powwellPercival1979JPhA}.

A \emph{dynamical map} associates a spectral number $N$ to each initial condition on a representative plane of initial conditions. The result is a  colour map, with a greyscale associated with $N$: Lighter regions represent regular orbits, while darker regions correspond to chaotic motion. The construction of the dynamical maps is complemented by calculating a \emph{dynamical power spectrum} along a family of initial conditions parameterized by one of the coordinates, for instance, the initial radial distance of the star.  The spectrum shows the evolution of the main frequencies of the problem and their linear combinations as functions of the chosen parameter.  The smooth evolution of frequencies is characteristic of regular motion, while the erratic scattering of the frequency values is characteristic of chaotic motion. The domains where one of the frequencies tends to zero accurately indicate the location of the separatrices between distinct regimes of motion, and resonance islands appear as regions between separatrices.
A detailed explanation of the numerical methods used in this paper is given in Appendix\,\ref{sec:SAM} \citep[see also][and references therein]{powwellPercival1979JPhA, michtchenkoEtal2002Icar, ferrazmeloMichtchenkoEtal2005LNP}.


\section{Equations of motion, stationary solutions, and spiral branches}\label{sec:spiralbranches}

The phase space of the Hamiltonian system under study is four-dimensional. The equations of motion of a star in the gravitational potential given by the Hamiltonian (\ref{eq1}) are written as

 \begin{equation}
\begin{array}{lrrrl}
\dfrac{dp_R}{dt} & = &-\dfrac{\partial{\mathcal H}}{\partial R}  &=& \dfrac{J_\varphi^2}{R^3}-\dfrac{\partial \Phi_0(R)}{\partial R} -\dfrac{\partial \Phi_1(R,\varphi)}{\partial R} ,\\
 & & & & \\
\dfrac{dR}{dt}   & = &\dfrac{\partial{\mathcal H}}{\partial p_R} &=& p_R ,\\
 & & & & \\
\dfrac{dJ_\varphi}{dt} & = &-\dfrac{\partial{\mathcal H}}{\partial \varphi} &=& -\dfrac{\partial \Phi_1(R,\varphi)}{\partial \varphi} , \\
 & & & & \\
\dfrac{d\varphi}{dt} & = &\dfrac{\partial{\mathcal H}}{\partial J_\varphi}  &=& \dfrac{J_\varphi}{R^2}-\Omega_p ,\\
\label{eq:eqsmotion}
\end{array}
\end{equation}
where \mbox{$\partial \Phi_0(R)/\partial R=V^2_{\rm rot}/R$}, as defined in Eq.\,(\ref{eq3}).

We look for stationary solutions of the Hamiltonian (\ref{eq1}), which correspond to circular orbits in the inertial frame rotating with angular velocity $\Omega_p$. The conditions for a star being at equilibrium in the rotating frame are \mbox{$dp_R/dt=dR/dt=dJ_\varphi/dt=d\varphi/dt=0$}.  The last three  conditions provide immediately that
\begin{eqnarray}\label{eq:H00}
 \nonumber 
  p_R &=& 0, \\
  m\,\varphi&=& \varphi_0 + f_m(R),\\
 \nonumber 
  J_\varphi &=& \Omega_pR^2 ,
\end{eqnarray}
where \mbox{$\varphi_0=\pm n\,\pi$} and \mbox{$n=0, 1, ...$}. The symmetry of the problem is $2\,\pi / m$, with the number of spiral arms $m$ equal to 2 in our case.

\begin{figure}
\begin{center}
\epsfig{figure=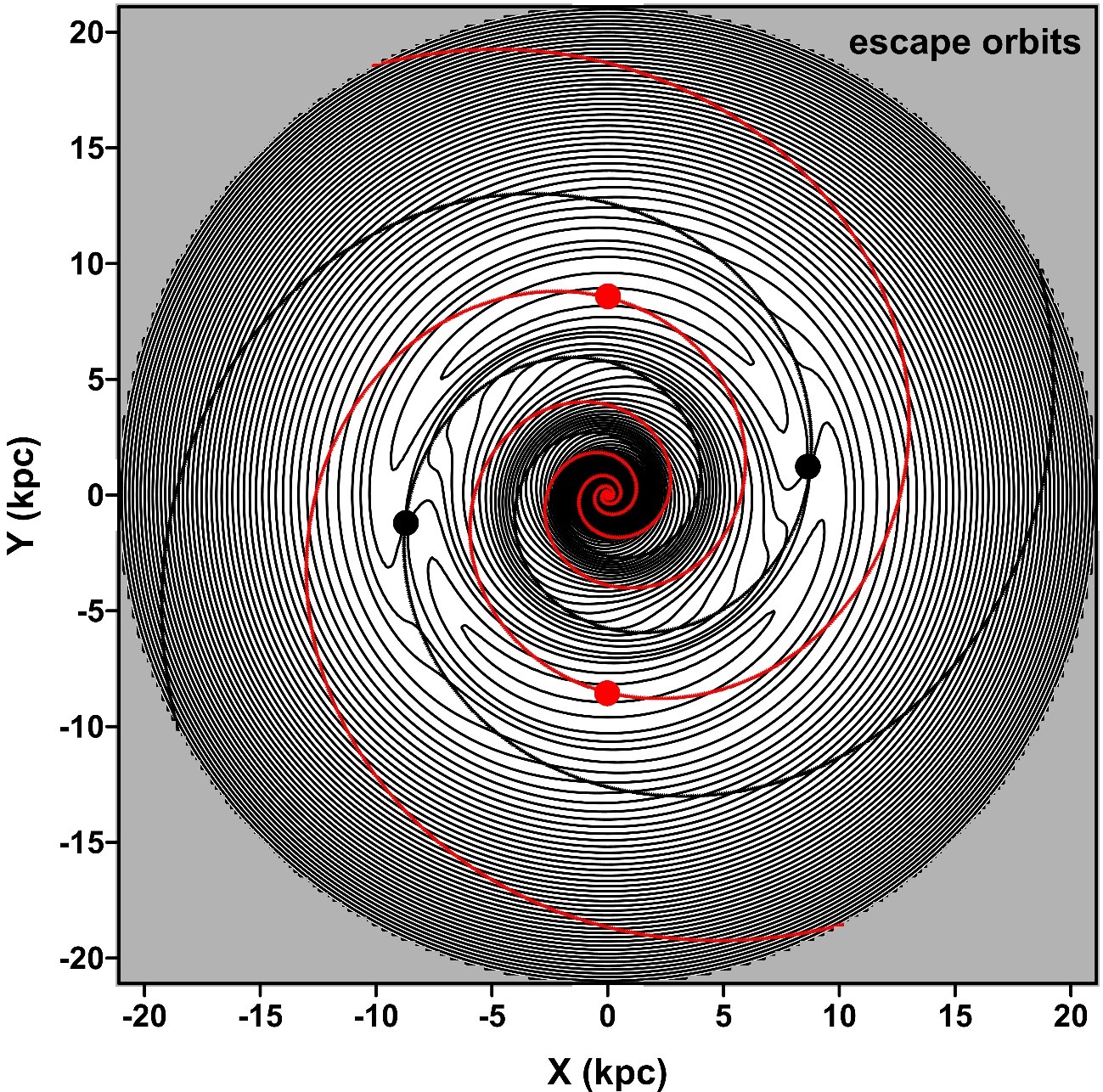,width=0.99\columnwidth ,angle=0}
\caption{Energy levels of the Hamiltonian function (\ref{eq1}) on the (\mbox{$X=R\cos\varphi$}, \mbox{$Y=R\sin\varphi$})--plane, for $p_r$ and $J_\varphi$ fixed at their stationary values given in Eq.\,(\ref{eq:H00}) and with the parameters from Table \ref{tab:1}, except \mbox{$\zeta_0=6300\,$km$^2$\,s$^{-2}$\,kpc$^{-1}$} and \mbox{$i=+14^\circ$}. Black lines correspond to the azimuthal minima of the Hamiltonian and are identified with the physical spiral arms. They correspond to the choice \mbox{$\varphi_0=0,\,2\pi$} in the second condition of Eq.\,(\ref{eq:H00}). Red lines correspond to azimuthal maxima of $\mathcal{H}$; they have \mbox{$\varphi_0=\pm\pi$} in the second condition of Eq.\,(\ref{eq:H00}). Dots represent the stationary solutions of the Hamiltonian system, elliptic (red) and hyperbolic (black). The grey region is of escaping orbits. The escape radius is obtained as \mbox{$R_{\rm esc}=21.16\,$kpc}.
}
\label{fig:contoursBracos}
\end{center}
\end{figure}

The conditions (\ref{eq:H00}) are visualized on the (\mbox{$X=R\cos\varphi$}, \mbox{$Y=R\sin\varphi$})--plane in Fig.\,\ref{fig:contoursBracos}, where they are plotted by the red and black lines, for \mbox{$\varphi_0=\pm\pi$} and \mbox{$\varphi_0=0, 2\pi$}, respectively. In the background of this figure, we plot energy levels of the Hamiltonian (\ref{eq1}) with $J_\varphi$ given by the last condition of Eq.\,(\ref{eq:H00}), calculated with the parameters from Table \ref{tab:1}, except \mbox{$\zeta_0=6300\,$km$^2$\,s$^{-2}$\,kpc$^{-1}$} and the pitch angle \mbox{$i=+14^\circ$}. We choose a larger value of $\zeta_0$  in order to enhance the visual effect of the perturbation. Also, the pitch angle $i$ is chosen positive in order to follow the conventional maps of the spiral structure of the Milky Way (e.g. \citealt[among others]{georgelinGeorgelin1976AA, drimmelSpergel2001ApJ, russeil2003AA, vallee2013IJAA, houHan2014AA}), with the Galactic rotation in the clockwise direction from the viewpoint of an observer located towards the direction of the north Galactic pole. All further calculations in this paper are done with the values of the parameters from Table\,\ref{tab:1}.

Figure \ref{fig:contoursBracos} shows that the Hamiltonian topology follows the black and red curves: the black branches correspond to the locations associated with the maxima of spiral arm density, while the red branches cross the level of maximum energy. These configurations satisfy the WKB approximation (for a tightly wound spiral pattern; see \citealp{binneytremaineGD}), which requires that \mbox{$\left|R\,\mathrm{d}f_{m}(R)/\mathrm{d}R\right|\gg 1$}, with $f_{m}(R)$ given in Eq.\,(\ref{eq6}); for our adopted parameters \mbox{$m=2$} and \mbox{$i=-14^\circ$}, we have \mbox{$\left|R\,\mathrm{d}f_{m}(R)/\mathrm{d}R\right|=\left|m/\tan(i)\right|\simeq 8$}. The stationary solutions of the Hamiltonian (\ref{eq1}) must belong to the branches  mentioned above, which we  refer to  hereafter  as \emph{spiral branches}.

The calculation of the solution for stationary $R$ is complicated and requires the implementation of some numerical procedures to resolve the condition \mbox{$dp_R/dt=0$} in Eq.\,(\ref{eq:eqsmotion}). The solutions obtained for equilibrium are shown by four large dots in Fig.\,\ref{fig:contoursBracos}. The equilibria belong to each one of the spiral branches. The two elliptic fixed points (red) correspond to the global maxima of the Jacobi constant $\mathcal{H}$ (\ref{eq1})
and lie on the red spiral branches, while the two hyperbolic saddle-like points (black) lie on the black spiral branches. It is worth noting that the stable solutions define the co-rotation radius. For the parameters from Table\,\ref{tab:1}, the co-rotation radius is approximately equal to $8.54$\,kpc and the phase angle is \mbox{$\varphi_{\rm eq} \cong -15^\circ.6$}. According to expression\,(\ref{eq6}), the free phase angle $\gamma$ initially fixed at zero, can now be assumed  to be equal to \mbox{$-m\,\varphi_{\rm eq}$}, placing the stable solutions always on the $Y$-axis, at \mbox{$\varphi=\pm 90^\circ$}.

\begin{figure}
\begin{center}
\epsfig{figure=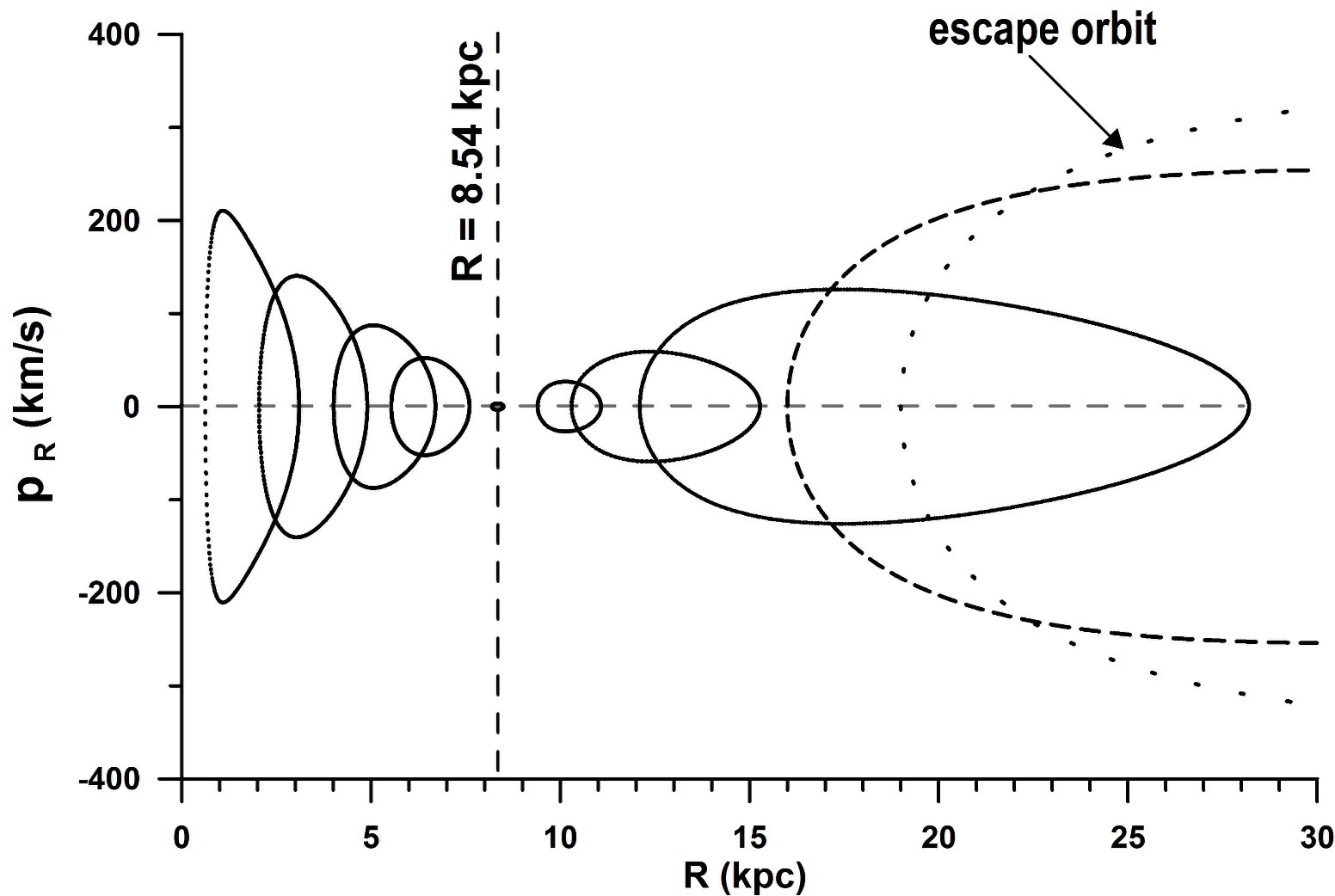,width=0.99\columnwidth ,angle=0}
\caption{Examples of stellar orbits calculated  with initial conditions along the spiral branch with \mbox{$\varphi_0= \pi$} (see Fig.\,\ref{fig:contoursBracos}).
}
\label{fig:orbitsPO}
\end{center}
\end{figure}

Figure \ref{fig:orbitsPO} shows stellar orbits  calculated for some initial conditions along a spiral branch defined by \mbox{$\varphi_0= \pi$}. The trajectories were  obtained by integrating numerically the equations of motion (\ref{eq:eqsmotion}) over only a few radial periods $T_R$; they are plotted in the phase subspace $R$--$p_R$. The location of the corotation radius is shown by the dashed line in Fig.\,\ref{fig:orbitsPO}. The evolution of the orbits starting on the spiral branch,  inside the co-rotation radius, is bound to this domain, and the amplitude of the $R$--oscillation grows when the initial conditions decrease from the co-rotation radius. The evolution of the spiral-branch orbits starting outside the co-rotation radius is analogous: they remain in this region and their amplitudes increase with increasing distance. At a given  distance, the orbits become unbounded; this distance defines the \emph{escape radius}, which is \mbox{$R_{\rm esc}\cong 21.16$\,kpc}, for the parameters from Table\,\ref{tab:1}. The shaded regions in Fig.\,\ref{fig:contoursBracos} are domains of initial conditions leading to escape orbits.

Outside the spiral branches, the escape velocity can be approximately obtained from the unperturbed problem in
Eq.\,(\ref{eq:Eunperturbed}) of Appendix\,\ref{sec:frequenciesUnperturbed},
at the condition \mbox{$E=0$}, as
\begin{equation}\label{eq:escapeboundaryV}
 V_{\rm esc}(R)=\sqrt{-2\Phi_0(R)}\,.
\end{equation}
This is a good approximation whenever the amplitude of the azimuthal perturbation is small. In the region surrounding the co-rotation point, the escape boundary (given by Eq.\,(\ref{eq:escapeboundaryV})) is around \mbox{$|V|= 600\,$\,km\,s$^{-1}$}.

\begin{figure}
\begin{center}
\epsfig{figure=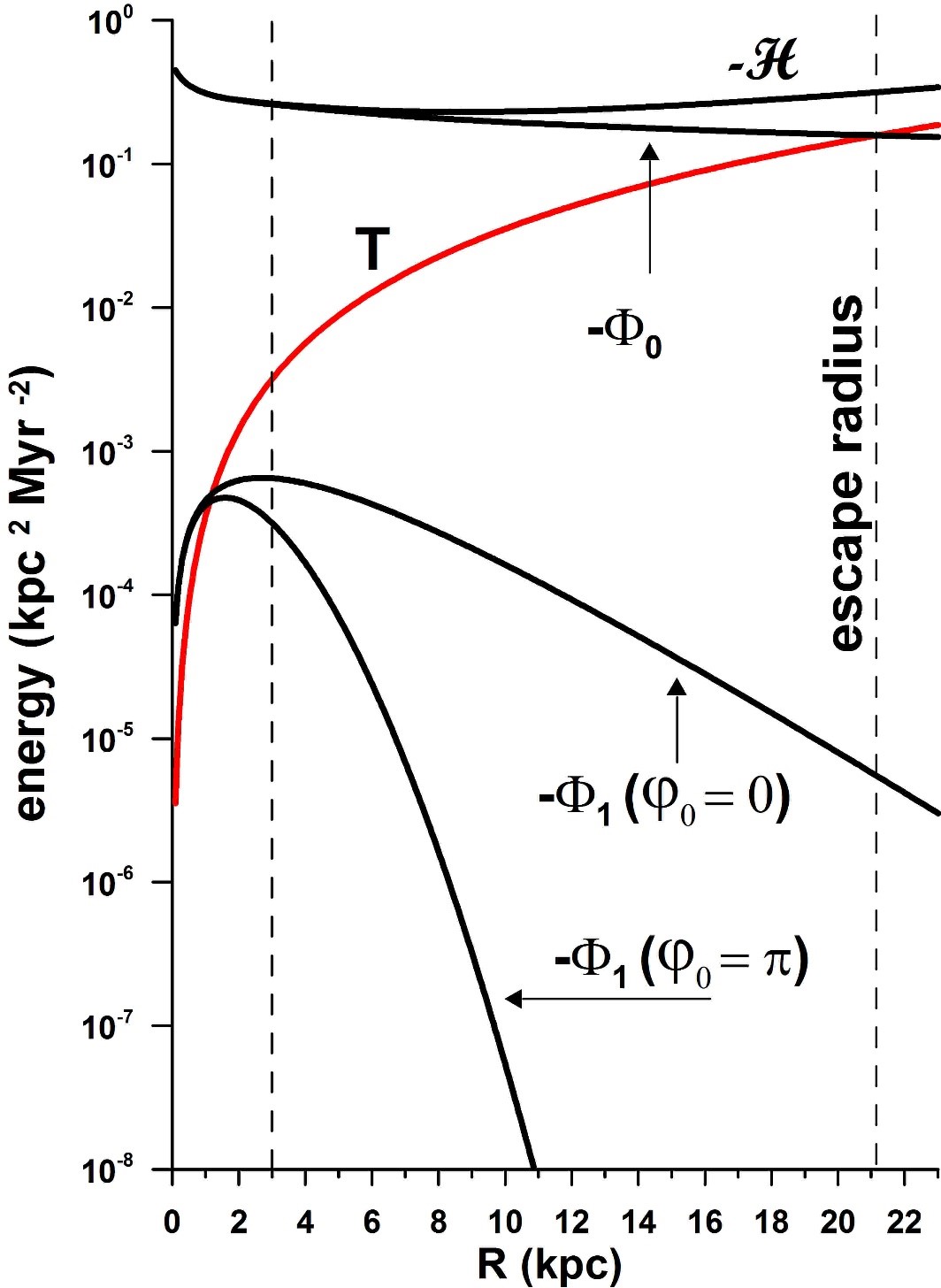,width=0.89\columnwidth ,angle=0}
\caption{Dependence of the energy components of Hamiltonian\,(\ref{eq1}) on the radius $R$, along the spiral branches (in logarithmic scale): $T$ is kinetic energy, $\Phi_0$ is the axially symmetric potential, and ${\mathcal H}$ is the Jacobi integral. Two spiral branches of the perturbation $\Phi_1$ are defined by \mbox{$\varphi_0 = 0$} and \mbox{$\varphi_0 = \pi$} (see Eq.\,(\ref{eq:H00})). The amplitude of the minima of the potential (\mbox{$\varphi_0=0$}) falls by two orders of magnitude from its maximum to the escape radius (outer vertical dashed line). The falling is almost exponential after \mbox{$R\cong 6\,$kpc}. We note that our model does not account for additional structures in the region below \mbox{$R=3$\,kpc} (inner dashed line), for instance, the Galaxy's bar.
}
\label{fig:energy}
\end{center}
\end{figure}

Figure \ref{fig:energy} shows the energy components of the Hamiltonian\,(\ref{eq1}), calculated along the spiral branches given by the conditions\,(\ref{eq:H00}), as functions of the radius $R$. We show the kinetic energy $T$, the axisymmetric potential \mbox{$\Phi_0$}, and the Jacobi integral $\mathcal{H}$. The radius value of $21.16$\,kpc, where the kinetic energy is equal to the modulus of the total gravitational potential energy \mbox{$\Phi_0+\Phi_1$}, defines the escape radius beyond which the spiral-branch stars are escaping.

It is interesting to observe in Fig.\,\ref{fig:energy} the evolution of the perturbation potential  $\Phi_1$ along the two spiral branches, one defined by \mbox{$\varphi_0=0$} and the other by \mbox{$\varphi_0=\pi$}. Both branches have maxima (of modulus) which lie close to 2\,kpc and are three orders of magnitude smaller than the Jacobi integral $\mathcal{H}$. For increasing distances, the strength of the perturbation decreases exponentially.


\section{Topology of $\mathcal{H}$ on the representative plane and stellar orbits}\label{topology}

To  visualize  the  dynamical  features  of  the  Hamiltonian system\,(\ref{eq1}),  we  introduce  a  representative  plane  of  initial conditions. The space of initial conditions of the two-degrees-of-freedom Hamiltonian system given by $\mathcal{H}$ is four-dimensional, but the problem can be reduced to the systematic study of initial conditions on the plane, which is a projection of a two-dimensional surface embedded in phase space. This plane may be chosen in such a way that all possible configurations of the system are included, and thus all possible regimes of motion of the system under study can be represented on it. Hereafter, we refer to this plane as a \emph{representative plane} of initial conditions.

In this work, we define the representative plane in the following way. First, we fix the initial values of the momentum $p_R$ at zero. Indeed, by definition, all bounded orbits should have two turning points defined by the condition \mbox{$p_R=0$}. In the following, we fix the initial values of the azimuthal angle $\varphi$. We know that this angle is generally circulating; it oscillates around $90^\circ$ (or $-90^\circ$) when the system is close to the stable stationary solution shown in Fig.\,\ref{fig:contoursBracos}. In both cases, it goes through $90^\circ$ (or $-90^\circ$) for all initial conditions. Hence, without loss of generality, the angular variable $\varphi$ can be initially fixed at $90^\circ$.

Now the topology of $\mathcal{H}$ can be fully represented on the plane of initial values $R$--$V_\theta$, where the stellar azimuthal velocity is defined as \mbox{$V_\theta=J_\varphi/R$}. We choose a different, non-canonical variable $V_\theta$ instead of $J_\varphi$ in order to have a better dynamical representation of orbits in terms of observables. The astrometric data from proper-motion measurements, along with the line-of-sight velocities and distances, can be transformed into $V_\theta$ data, allowing us to develop a framework in which future observations can be fit in.

It is worth emphasizing that the $X$--$Y$ plane widely used in the stellar dynamics studies (see Fig.\,\ref{fig:contoursBracos}) cannot be chosen as a representative plane since, in its construction, the initial values of the angular momentum are fixed at the condition \mbox{$J_\varphi = \Omega_pR^2$}.  The Hamiltonian topology, in this case restricted to the stationary values of $p_r$ and $J_\varphi$, is equivalent to the analysis of the  level curves of the effective potential \mbox{$\Phi_{\rm eff}=\Phi_0+\Phi_1-\Omega_p R^2/2$}.
The energy function is written as \mbox{$h(R,\varphi,\dot{R},\dot{\varphi})= \mathcal{H}(R,\varphi,p_R,J_\varphi)$}, where
\mbox{$h=\big[\dot{R}^2+R^2\dot{\varphi}^2\big]/2 + \Phi_{\rm eff}(R,\varphi)$}
(see e.g. \citealp{binneytremaineGD,barrosLepineJunqueira2013MNRAS}). The equations of motion are \mbox{$\mathbf{\ddot{x}}=-\nabla\Phi_{\rm eff} -2\mathbf{\Omega}_p\times\mathbf{\dot{x}}$}, with \mbox{$\mathbf{\Omega}_p=\Omega_p\mathbf{\hat{z}}$}.

\begin{figure*}
\begin{center}
\epsfig{figure=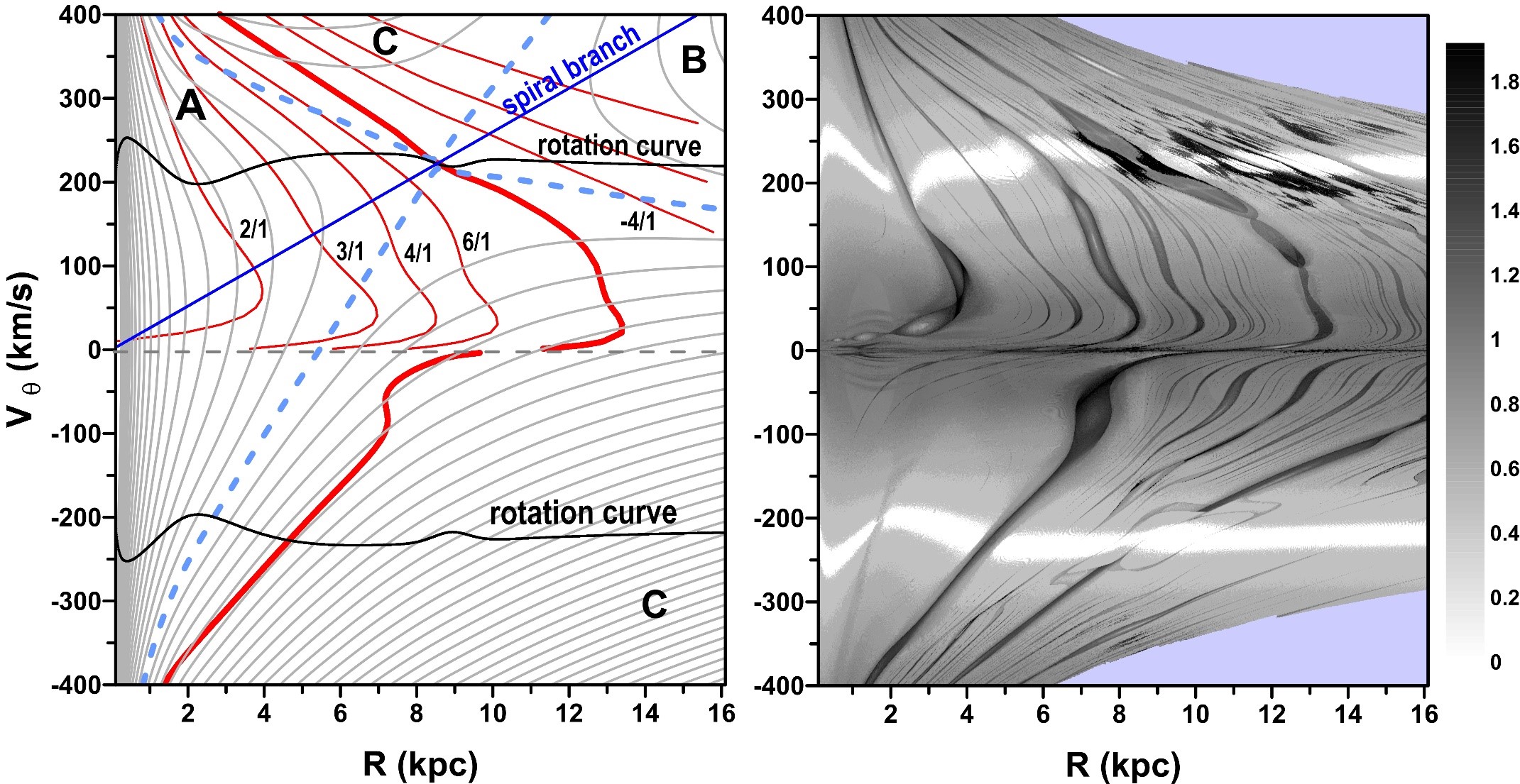,width=0.99\textwidth ,angle=0}\\
\caption{Left: Topology of $\mathcal{H}$ (\ref{eq1}) on the $R$--$V_\theta$ plane of initial conditions, where \mbox{$V_\theta=J_\varphi/R$}, \mbox{$p_R=0$}, and \mbox{$\varphi=90^\circ$}. Right: Dynamical maps on the same plane calculated for the time series $J_\varphi(t)$. The shaded blue regions correspond to domains of initial conditions leading to orbits which go beyond 30\,kpc (the region where our model is no longer applicable). The bar relates grey tones and the values of the spectral number $N$ between 1 and 100 in logarithmic scale. All orbits with \mbox{$N>100$} are labelled in black. A typical orbit intersects this plane in two points corresponding to the minimum and maximum values of the coordinate $R$.
}
\label{fig:VRtopologyANDmap}
\end{center}
\end{figure*}

Figure \ref{fig:VRtopologyANDmap} shows, in the left panel, the energy levels of the Hamiltonian on the plane $R$--$V_\theta$ (solid grey lines). The level which contains the stationary solution with coordinates \mbox{$R=8.54$\,kpc} and \mbox{$V_\theta=221.8$\,km\,s$^{-1}$} is shown by a dashed blue line. It separates  the whole domain in three dynamically distinct regions, \textbf{A}, \textbf{B}, and \textbf{C}, which is explained below. The  straight blue line shows the \mbox{($R,\,V_\theta$)}  coordinates of the spiral branches.

The black curves present locations of  the rotation curve $V_{\rm rot}$\,(\ref{eq:Vrot}) for both positive and negative values of $V_\theta$. In the unperturbed problem, when $J_\varphi$ remains constant in time, the rotation curves define the locations of circular orbits according to Eqs.\,(\ref{eq:eqsmotion}), while the initial conditions outside the rotation curves lead to oscillations around the circular orbits with the corresponding values of $J_\varphi$. Hereafter we  refer  to this pattern of oscillation as a \emph{$R$--mode of motion}. The orbits starting with \mbox{$V_\theta>0$} (\mbox{$V_\theta<0$}) are \emph{prograde (retrograde) orbits} and oscillate around the positive (negative) branch of the rotation curve.

When the perturbation due to spiral arms is introduced in the problem, the $R$--mode  remains as a dominating mode of motion, while $J_\varphi$ begins  oscillating according to Eqs.\,(\ref{eq:eqsmotion}). We  refer to this new pattern of oscillation as a \emph{$J_\varphi$--mode of motion}. Owing to this additional mode of motion, the  circular orbits of the unperturbed problem become non-circular. The initial conditions, for which the amplitude of the $R$--mode tends to zero, form a family of periodic orbits; if the perturbation is small, the periodic orbits are nearly circular orbits and their family is located in the vicinity of the positive and negative branches of the rotation curve, depending on the direction of the initial $V_\theta$--value. Therefore, the perturbed system oscillates in the $R$--mode around non-circular periodic orbits, with the exception of the equilibrium points.

Thus, a typical stellar motion in the full phase space can be represented as a combination of two independent oscillations, determined by the $R$-- and $J_\varphi$--modes of motion  with characteristic frequencies $f_R$ and $f_{\varphi}$, respectively.

\begin{figure}
\begin{center}
\epsfig{figure=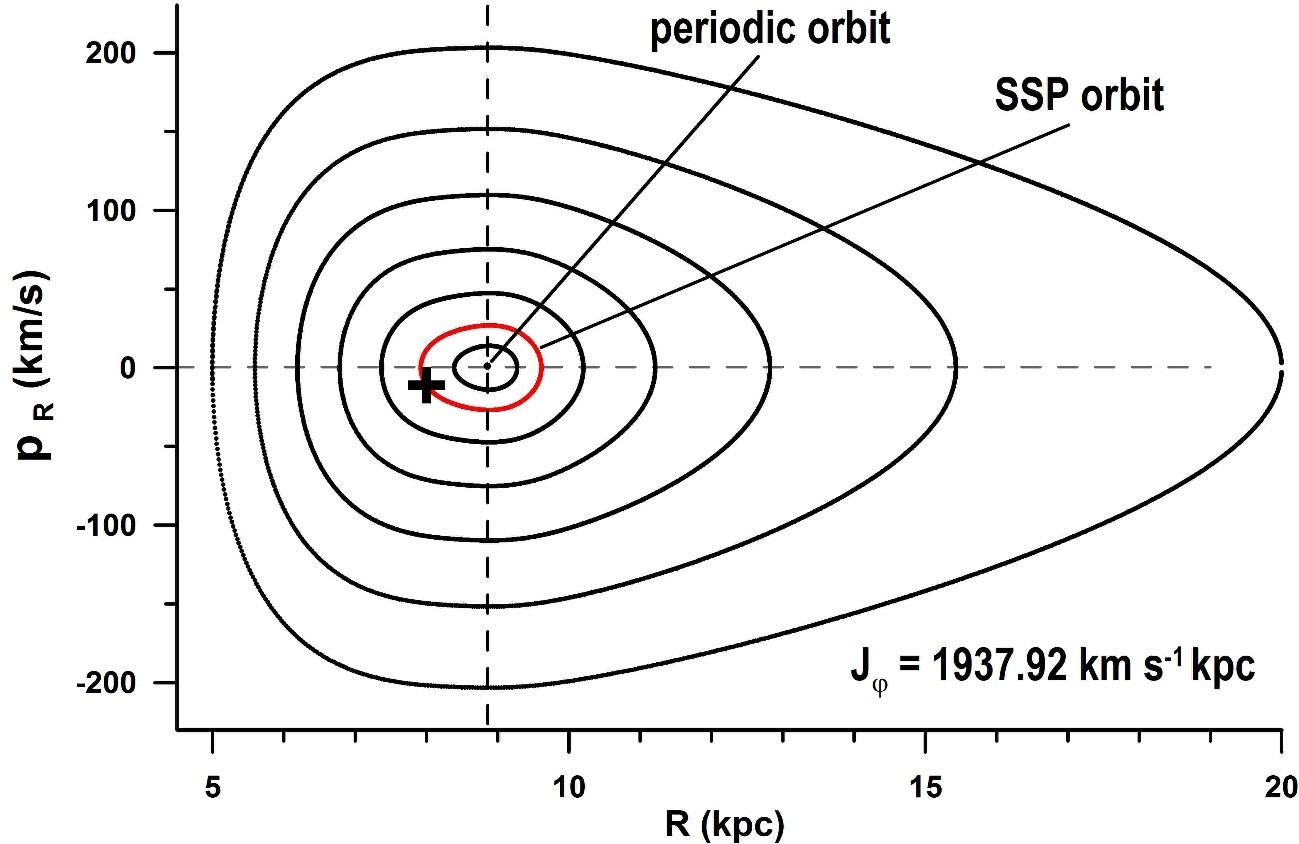,width=0.99\columnwidth ,angle=0}
\caption{Surfaces of section on the $R$--$p_R$ plane calculated with \mbox{$J_\varphi=1937.92$\,km\,s$^{-1}$\,kpc}, which corresponds to the current parameters of the SSP (see text). The orbit of the SSP and its current position are shown by a red curve and a cross symbol, respectively.
}
\label{fig:SOSsun}
\end{center}
\end{figure}

To illustrate the behaviour of a star in the $R$--mode, we show in Fig.\,\ref{fig:SOSsun} the surfaces of section in the phase subspace \mbox{$R$--$p_R$} calculated along the fixed value of \mbox{$J_\varphi=1937.92$\,km\,s$^{-1}$\,kpc}. This value corresponds to the current coordinates of the Sun, \mbox{$R_{\odot}=R_0=8$\,kpc} and \mbox{$V_{\theta,\odot} = V_{0}+v_{\odot} = 242.24$\,km\,s$^{-1}$} ($V_0$ and $v_{\odot}$ were defined in Sect.\,\ref{sec:rotcurve}). The family of orbits is centred around a periodic orbit, with zero-amplitude $R$--mode oscillation, corresponding to the chosen value of $J_\varphi$. The energy of the periodic orbit is minimal for a given $J_\varphi$, and increases when the oscillation amplitude  increases. Knowing the current value of the Sun's linear momentum, \mbox{$p_R=-11$\,km\,s$^{-1}$}, we plot the orbit of the star with solar kinematic parameters (SSP) and its position by the red curve and a cross symbol, respectively, in Fig.\,\ref{fig:SOSsun}. The period of the $R$--mode oscillation of the SSP (inverse of $f_R$) is approximately 190\,Myr.  It is worth noting that, for small perturbations when \mbox{$J_\varphi\approx const$},  the described behaviour in the phase subspace $R$--$p_R$ is determined mainly by the unperturbed term ${\mathcal H_0}$ in the Hamiltonian (\ref{eq1}).

The typical behaviour of a star under the perturbation due to spiral arms is illustrated in the phase subspace \mbox{$\varphi$--$V_\theta$} (the azimuthal velocity \mbox{$V_\theta=J_\varphi/R$}) in Fig.\,\ref{fig:sos-all}. The top panel shows the surface of section for prograde orbits calculated along the energy level \mbox{${\mathcal H}=-0.2298$\,kpc$^2$\,Myr$^{-2}$} close to the SSP energy. The bottom panel
shows the surface of section for retrograde orbits along the energy level \mbox{${\mathcal H}=-0.20$\,kpc$^2$\,Myr$^{-2}$}. The conditions used in the construction of the surfaces of section were \mbox{$p_R=0$} and \mbox{$\dot{p}_R<0$}, which reduce the four-dimensional phase space of the full problem to a two-dimensional surface defined by the variables $\varphi$ and $V_\theta$. We associate the behaviour of the system shown in Fig.\,\ref{fig:sos-all} with the $J_\varphi$-- mode of oscillation of the system, with the corresponding frequency $f_\varphi$.

Since the perturbation determined by the term $\Phi_1$ in the Hamiltonian\,(\ref{eq1}) is small (see Fig.\,\ref{fig:energy}), the variation of the azimuthal velocity is weak. It is enhanced only by the resonances which appear as chains of islands; some of the resonances are identified  in Fig.\,\ref{fig:sos-all}, top. The most significant of these is the \emph{co-rotation resonance}, which occurs at \mbox{$V_\theta > 0$}, when the azimuthal frequency  $f_\varphi$ defined in Eqs.\,(\ref{eq:fphifR}) tends to zero. The estimate of the current $\varphi$--value of the SSP is discussed in Sect.\,\ref{sec:solar}. The 1/1 resonance observed in Fig.\,\ref{fig:sos-all}, bottom, for retrograde orbits with \mbox{$V_\theta < 0$} is discussed in the next section.

Finally, the analysis of the topology of the Hamiltonian, shown in Fig.\,\ref{fig:VRtopologyANDmap}, left, allows us to infer  several constraints on the stellar orbital evolution under the potential of the model galaxy. For instance, the orbits starting in the domains \textbf{A} and \textbf{B}  are characterized by energies below the equilibrium value. Since the Jacobi integral ${\mathcal H}$\,(\ref{eq1}) is conserved during the orbital evolution of the star, all initial conditions from region \textbf{A} lead to either prograde or retrograde orbits (apart from the resonant regions around \mbox{$V_\theta=0$}), which are all confined inside this region, with maximum distance smaller than the co-rotation radius. In contrast, all stars starting in the region \textbf{B} have prograde orbits and evolve beyond the co-rotation radius. We note that spiral branches, as well as prograde orbits starting along the rotation curve, belong to  regions \textbf{A} and \textbf{B}.

In region \textbf{C} of Fig.\,\ref{fig:VRtopologyANDmap}, left, the stellar orbits have energies higher than the equilibrium energy: the amplitudes of the \mbox{$R$--mode} of oscillation are small when the star starts close to the families of periodic solutions and rapidly increase with increasing deviation from these families. The only constraint is that prograde orbits remain prograde forever, and also that retrograde orbits remain retrograde (apart from the resonant regions around \mbox{$V_\theta=0$}), since the variation of $J_\varphi$ is very small under small perturbations. In particular, the inner turning point $R_1$ will have in general \mbox{$|V_\theta(R_1)|>|V_{\rm rot}(R_1)|$} and the outer turning point $R_2$ will have \mbox{$|V_\theta(R_2)|<|V_{\rm rot}(R_2)|$} for both prograde and retrograde motion. This region is unbounded for \mbox{|$V_\theta|\to\pm\infty$}.

\section{Dynamical maps, dynamical power spectra, and resonances}\label{sec:maps}

The dynamical map of the representative plane $R$--$V_\theta$ is shown in Fig.\,\ref{fig:VRtopologyANDmap}, right. During its construction, for each initial condition of the 1000$\times$1000-grid, the equations of motion\,(\ref{eq:eqsmotion}) were integrated numerically over \mbox{$\sim 2600$} radial epicyclic periods, with the value of $R$ given by the initial conditions of the orbit. The values of $p_R$ and $\varphi$ are fixed at 0 and $90^\circ$, respectively. Next, the time series of the variable $J_\varphi$ were SAM--analysed and their spectral numbers $N$ determined. The spectral numbers were then plotted on the representative plane using a greyscale code  in logarithmic scale (see colour bar in Fig.\,\ref{fig:VRtopologyANDmap}, right). This scale varies from 0 to 2, corresponding to a spectral number $N$ between 1 and 100. All values \mbox{$N>100$} are also labelled in black.

All initial conditions leading to orbits whose radius exceeds 30\,kpc at some moment during the integrations were plotted in blue in Fig.\,\ref{fig:VRtopologyANDmap}, right. This value was chosen because several massive structures, such as the Sagittarius dwarf spheroidal galaxy \citep{helmi2004MNRAS,lawEtal2009ApJL,degWidrow2013MNRAS,ibataEtal2013ApJL} and the Magellanic clouds among other possible candidates,  orbit our Galaxy at approximately  this distance.
Our model does not take into account interactions with these structures and therefore is no longer valid  at regions beyond 30\,kpc.

Comparing the two graphs in Fig.\,\ref{fig:VRtopologyANDmap}, we note the correlation between the white horizontal strips on the dynamical map (right panel) and the location of the rotation curves in the left panel, for both prograde and retrograde orbits. Indeed, the colour white  in the greyscale corresponds to harmonic motion, characterized by a very small value of the spectral number $N$. This is also an important property of the periodic solutions, for which the amplitude of the $R$--mode of oscillation tends to zero. We note that there is no family of periodic solutions composed of orbits with zero-amplitude $J_\varphi$--mode  of oscillations.

\begin{figure}
\begin{center}
\epsfig{figure=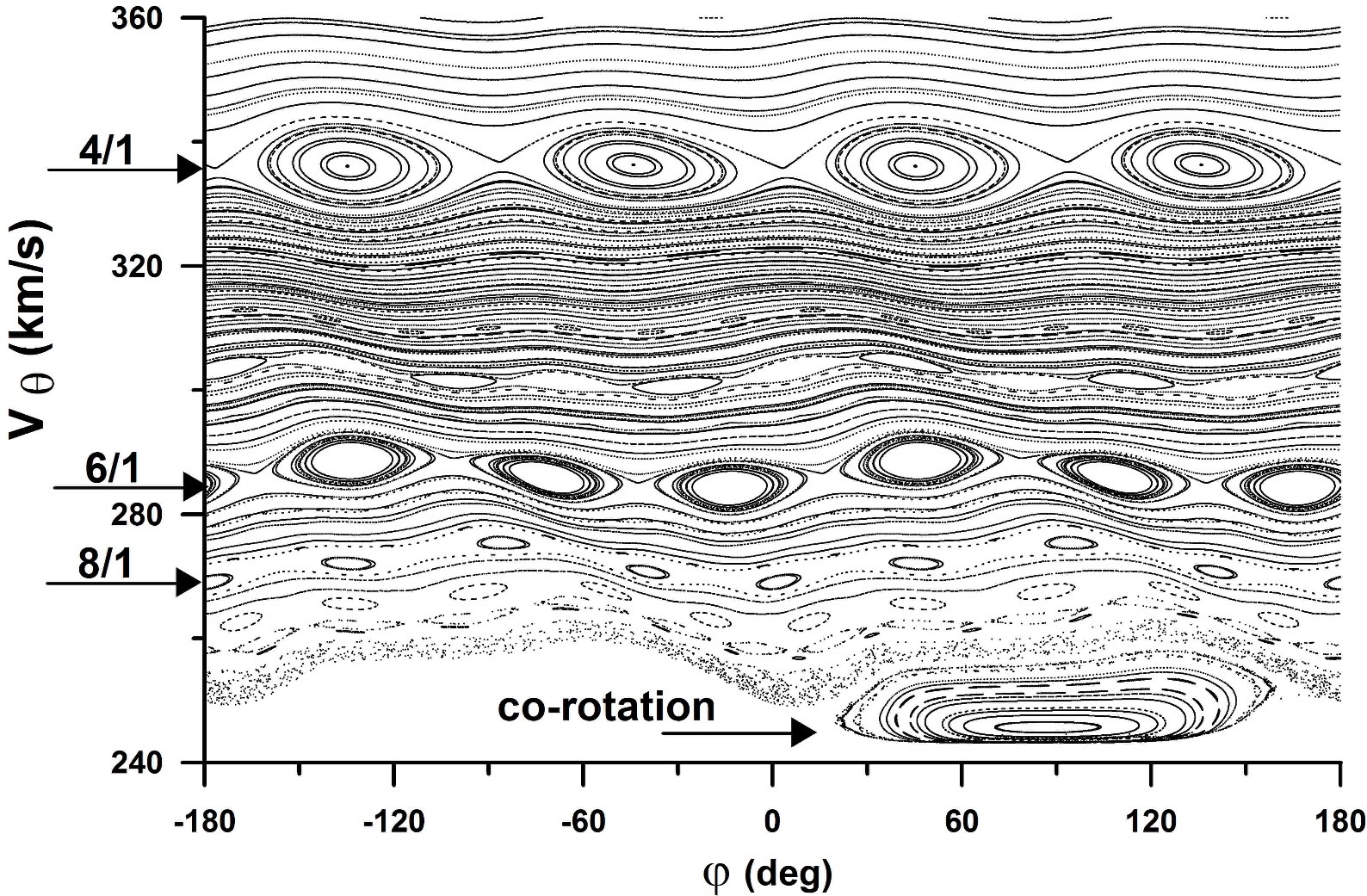,width=0.99\columnwidth ,angle=0}\\
\vspace{0.5cm}
\epsfig{figure=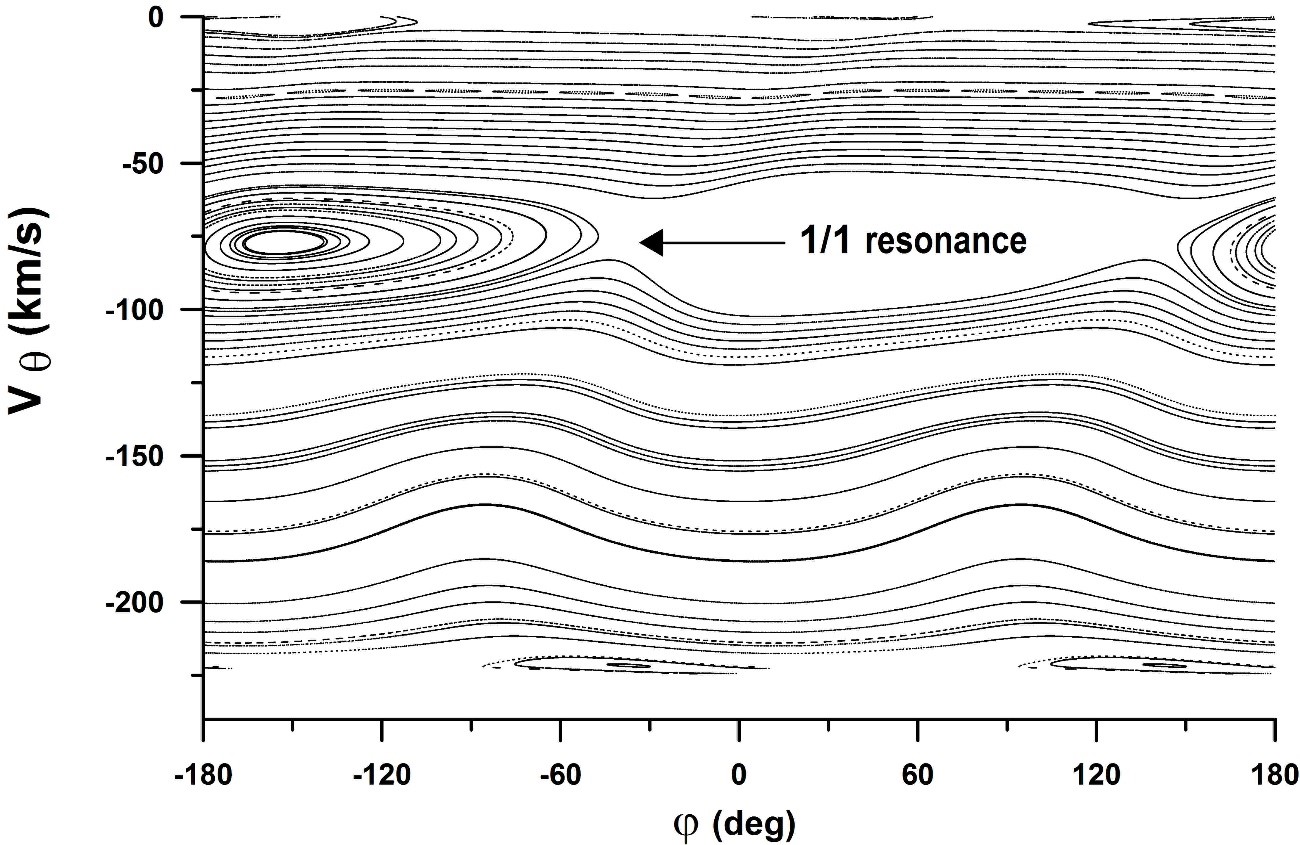,width=0.99\columnwidth ,angle=0}
\caption{Top: Surfaces of section on the $\varphi$--$V_\theta$ plane calculated along the energy level \mbox{${\mathcal H}=-0.2298$\,kpc$^2$\,Myr$^{-2}$}, close to the SSP energy, and \mbox{$V_\theta>0$}. Several resonances appear as chains of islands (the number of islands is equal to the order of the resonance); some are indicated. The estimated values of $\varphi$ for the SSP (52$^\circ$ or 142$^\circ$, see Sect.\,\ref{sec:solar}) place it inside the co-rotation resonance (see also Fig.\,\ref{fig:MapParametersSSP}); the period of the $J_\varphi$--mode of oscillation of the SSP is about 2\,Gyr. Bottom: Same as the top panel, except that \mbox{${\mathcal H}=-0.20$\,kpc$^2$\,Myr$^{-2}$} and \mbox{$V_\theta<0$}.
}
\label{fig:sos-all}
\end{center}
\end{figure}

On the contrary, the narrow black strips observed on the dynamical map in Fig.\,\ref{fig:VRtopologyANDmap}, right, are characterized by highly non-harmonic and chaotic stellar motions. They are associated with the dynamical phenomenon known as a \emph{resonance}. A resonance occurs when one of the fundamental frequencies of the system or one of the linear combinations of these frequencies tends to zero. In this case, the topology of the phase space is transformed, giving rise to islands of stable resonant motion which are surrounded by the layers of chaotic motion associated with the separatrix of the resonance \citep{Ferraz-Mello2007}. Some examples of the formation of the chains of islands and the libration of the angle $\varphi$ can be observed in Fig.\,\ref{fig:sos-all}.

\begin{figure*}
\begin{center}
\epsfig{figure=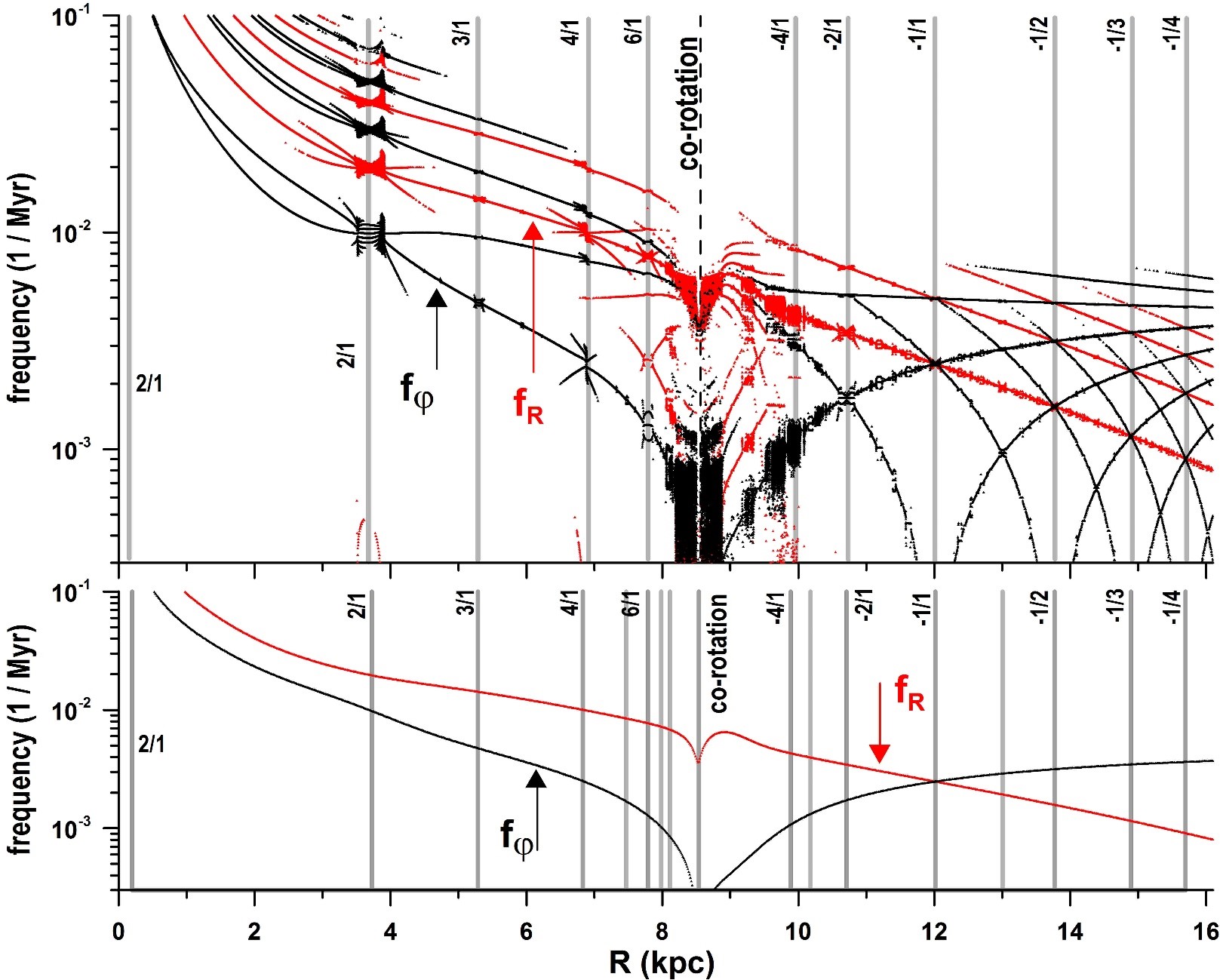,width=0.94\textwidth ,angle=0}\vspace{0.3cm}
\epsfig{figure=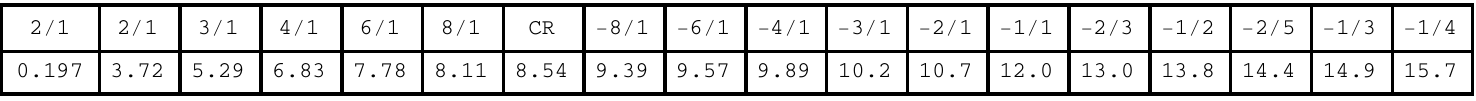,width=0.99\textwidth ,angle=0}
\caption{Top: Dynamical power spectrum: the evolution of the proper frequencies $f_R$ (red) and $f_\varphi$ (black), their harmonics, and linear combinations along the spiral branch with \mbox{$\varphi_0=\pi$}, in logarithmic scale. The frequencies $f_R$ and $f_\varphi$ were calculated by analysing the time evolution of $R$ and $J_\varphi$.  The smooth evolution of the frequencies is characteristic of regular motion, while the erratic scattering of the points is characteristic of chaotic motion (see Sect.\,\ref{sec:tools} and Appendix\,\ref{app2}). The main Lindblad resonances (see Sect.\,\ref{sec:epicyclic}) are indicated by vertical lines and the corresponding ratio. The sign `-' corresponds to the condition \mbox{$\Delta\varphi < 0$} in Eq.\,(\ref{eq:Deltavarphi}) (see also Appendix\,\ref{sec:frequenciesUnperturbed}).
Bottom: Frequencies $f_R$ (red) and $f_\varphi$ (black) calculated via the equations in Appendix \ref{sec:frequenciesUnperturbed}, along the curve \mbox{$J_\varphi(R)=\Omega_p\,R^2$}, with the unperturbed potential $\Phi_0$\,(\ref{eq:H0}). They match the fundamental frequencies of the perturbed problem (top panel), and the resonance radii (grey vertical lines) are very close to the corresponding radii of the perturbed problem.
The bottom table shows the radii of resonances in the unperturbed problem.
}
\label{fig:specJ1Omegap}
\end{center}
\end{figure*}

The identification of the resonances observed on the dynamical map in Fig.\,\ref{fig:VRtopologyANDmap}, right, is done by applying the dynamical power spectrum approach described in Appendix\,\ref{sec:SAM}. The dynamical spectrum obtained shows the evolution of the proper frequencies, $f_R$ (red) and $f_\varphi$ (black), and their linear combinations along the spiral branch with \mbox{$\varphi_0=\pi$}, in Fig.\,\ref{fig:specJ1Omegap}, top. The dominant feature of the spectrum is the  trend of the frequency $f_\varphi$ associated with the $J_\varphi$--mode of motion towards zero-values at distances close to the co-rotation radius, around $8.54$\,kpc (vertical dashed line). We associate this behaviour with the co-rotation resonance which occurs at the condition \mbox{$\bar\Omega_\theta \cong \Omega_p$}, where $\bar\Omega_\theta$ is the averaged value of the angular velocity of the star along its trajectory.  It is worth emphasizing here that, at the co-rotation radius, this resonance is crossed by the family of periodic solutions with \mbox{$V_\theta >0$}; thus, the intersection of the spiral branch with the family of periodic solutions gives rise to a stationary configuration of the system.   The dynamical map constructed with $\varphi=90^\circ$ (Fig.\,\ref{fig:VRtopologyANDmap}, right) shows a wide stable region around  the co-rotation point, surrounded by layers of highly unstable motion.

The co-rotation resonance appears in Fig.\,\ref{fig:VRtopologyANDmap}, right, as the black strip emanating from the stationary solution and crossing the positive half-plane $R$--$V_\theta$. The location of the co-rotation resonance can be estimated analytically through the unperturbed approximation assuming \mbox{$\Delta\varphi=0$} in Eq.\,(\ref{eq:Deltavarphi}) (see also Appendix\,\ref{sec:frequenciesUnperturbed}). Figure~\ref{fig:specJ1Omegap}, bottom,  shows the fundamental frequencies $f_R$ (red) and $f_\varphi$ (black) calculated via  Eq.\,(\ref{eq:Deltavarphi}) and the equations in Appendix\,\ref{sec:frequenciesUnperturbed}, with the unperturbed potential $\Phi_0$\,(\ref{eq:H0}). The comparison between the fundamental frequencies obtained analytically and those of the perturbed problem obtained numerically shows a good qualitative agreement, at least when the perturbations are small. Obtained  this way, the approximate location of the co-rotation resonance on the representative plane $R$--$V_\theta$ is shown by a thick red curve on the positive half-plane in Fig.\,\ref{fig:VRtopologyANDmap}, left.

In addition to the co-rotation resonance, there is another kind of resonance whose effects can be  observed in Fig.\,\ref{fig:specJ1Omegap}, top, as discontinuities in the smooth evolution of the frequencies in the dynamical power spectrum. These discontinuities are associated with  the initial configurations for which the two fundamental frequencies of the problem become commensurable; in other words, they obey the relation \mbox{$\textrm{j}\,f_R - \textrm{n}\,f_\varphi \cong 0 $}, where $\textrm{j}$ and $\textrm{n}$ are simple integers. We say that, when the above condition is satisfied, a resonance occurs. We have identified these resonances and marked their positions by vertical lines and the corresponding ratio $\textrm{n}/\textrm{j}$ in Fig.\,\ref{fig:specJ1Omegap}. The positions of the resonances, obtained for the unperturbed problem via Eq.\,(\ref{eq:Deltavarphi}) and Appendix\,\ref{sec:frequenciesUnperturbed}, are also shown in the bottom panel in Fig.\,\ref{fig:specJ1Omegap}, together  with the approximated values of the resonance radii. It is worth noting a very good agreement between the results obtained analytically for the axisymmetric problem and numerically for the full problem.

\subsection{Comparison with the epicyclic approximation}\label{sec:epicyclic}

It is known that the epicyclic approximation allows us to predict the positions of resonances along nearly circular orbits which have small radial deviations \citep[and references therein]{lindblad1974IAUS, contopoulosGrosbol1986AA, binneytremaineGD, efthymiopoulos2010EPJST}.
Those studies gave rise to the concept of \emph{Lindblad resonances}  (e.g.  \citealp{binneytremaineGD}). Some recent studies base the interpretation of numerical results on the positions of resonances predicted by the epicyclic approximation (e.g. \citealp[among others]{dehnen2000AJ,quillen2003AJ,quillenMinchev2005AJ, minchevQuillen2006MNRAS,antojaEtal2011MNRAS,gomezEtal2013MNRAS,
barrosLepineJunqueira2013MNRAS,faureSiebertFamaey2014MNRAS}).

\begin{figure*}
\begin{center}
\epsfig{figure=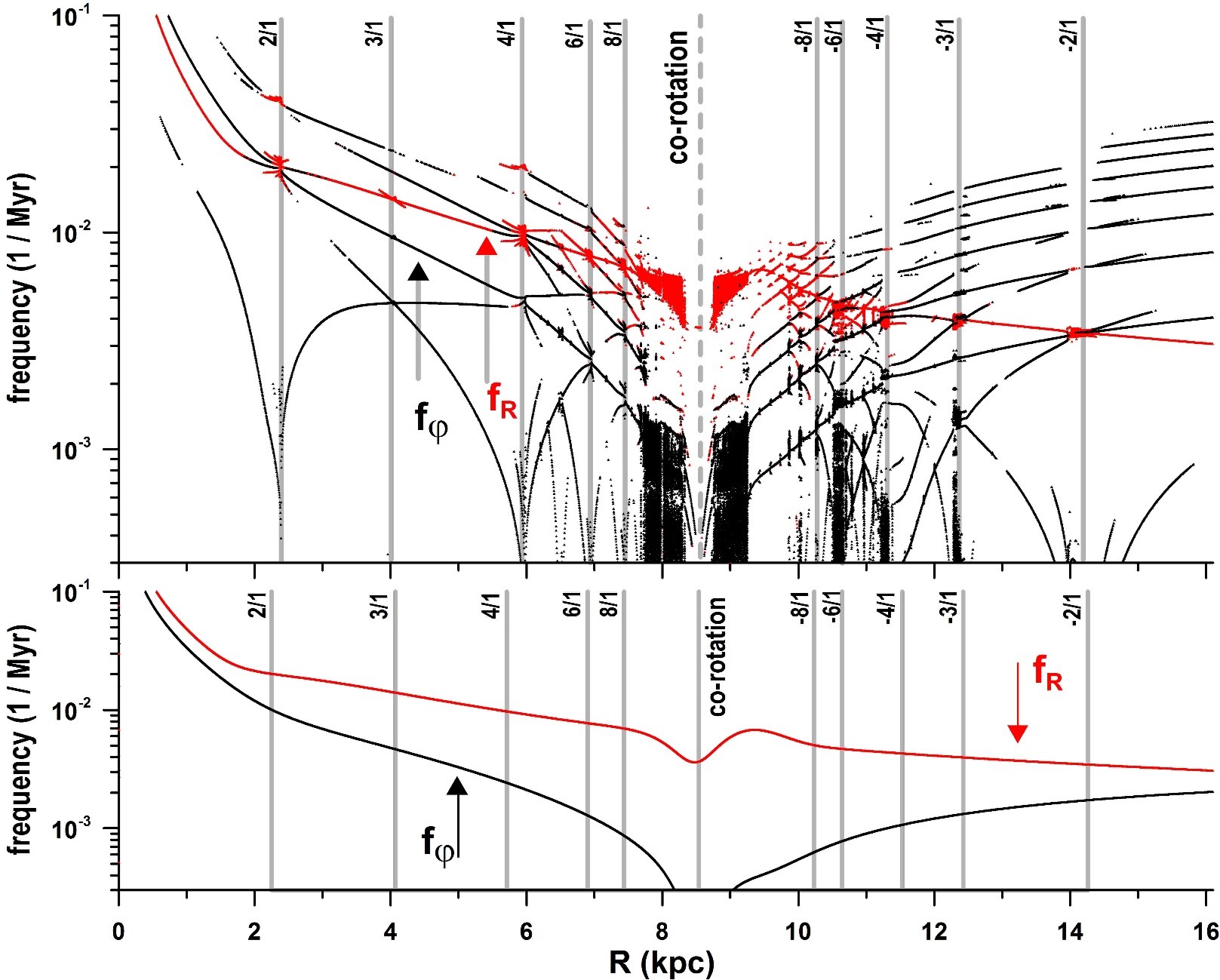,width=0.94\textwidth ,angle=0}\vspace{0.3cm}
\epsfig{figure=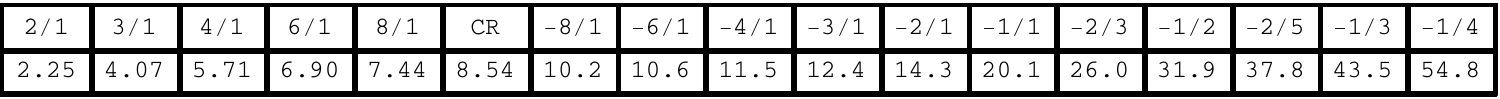,width=0.99\textwidth ,angle=0}
\caption{Same as in Fig.\,\ref{fig:specJ1Omegap}, except that the frequencies were calculated along the rotation curve\,(\ref{eq:Vrot}) with fixed \mbox{$\varphi=90^\circ$}. The epicyclic approximation was employed in order to estimate the frequencies analytically.
}
\label{fig:specCircular}
\end{center}
\end{figure*}

In our approach, we expand the definition of these resonances to the whole representative domain of stellar orbits; moreover, since these newly determined resonances are of the same nature, we continue to refer to them as Lindblad resonances. We show in Fig.\,\ref{fig:specCircular} the dynamical power spectrum of the orbits along \mbox{$V_\theta=V_{\rm rot}(R)$}, with \mbox{$\varphi=90^\circ$}. The top panel shows the full dynamical power spectrum, calculated for the perturbed Hamiltonian. The bottom panel shows the unperturbed prediction for the radial and azimuthal frequencies $f_R$ and $f_{\varphi}$, and the corresponding predictions for the resonance radii.  We see that the fundamental frequencies associated with the $R$-- and $J_\varphi$--modes agree with those obtained in the unperturbed case (compare the top and bottom panels of Fig.\,\ref{fig:specCircular}). The radii of the Lindblad (epicyclic) resonances match the unperturbed prediction, Eq.\,(\ref{eq:LindbladRess}), the difference being due to the spiral perturbation.

The difference between the epicyclic (Lindblad) approximation (Fig.\,\ref{fig:specCircular}) and the resonances of the dynamical power spectrum calculated along the spiral branch (Fig.\,\ref{fig:specJ1Omegap}) are  due to the adopted form of the curve $V_\theta(R)$ for the initial conditions on the \mbox{$R$--$V_\theta$} dynamical map of Fig.\,\ref{fig:VRtopologyANDmap}. The epicyclic approximation considers $V_\theta$ given by circular motion in the unperturbed potential, \mbox{$V_\theta(R)=V_{\rm rot}(R)$}.

Finally, the prediction for the Lindblad resonance locations (in our context)  on the representative plane \mbox{$R$--$V_\theta$}, are shown in the left panel in Fig.\,\ref{fig:VRtopologyANDmap} by red curves and the corresponding ratio $\textrm{n}/\textrm{j}$.
These predictions were obtained from the unperturbed problem via Eq.\,(\ref{eq:Deltavarphi}) and the equations in Appendix \ref{sec:frequenciesUnperturbed}.
The corresponding epicyclic resonances are precisely the intersections between quasi-circular orbits (represented by a white strip around the rotation curve in Fig.\,\ref{fig:VRtopologyANDmap}, right) and the resonance chains, and their predictions from the unperturbed problem are given by the intersection between the rotation curve and the red curves in Fig.\,\ref{fig:VRtopologyANDmap}, left.

\subsection{Retrograde orbits}\label{sec:retrograde}

The dynamical portrait of the domain of retrograde orbits can be seen on the negative half-plane in Fig.\,\ref{fig:VRtopologyANDmap}. According to the solutions for the extrema of the Hamiltonian function\,(\ref{eq1}), there is no co-rotation point in
this region (and, consequently, no co-rotation resonance), which is dominated by the strong 1/1 Lindblad resonance. The position of this resonance can be predicted from the unperturbed potential  $\Phi_0$ analytically, assuming \mbox{$\Delta\varphi=0$} in Eq.\,(\ref{eq:Deltavarphi}) and noting that, for \mbox{$V_\theta<0$}, we have \mbox{$\Delta\theta<0$} in Eq.~(\ref{eq:DeltaThetaaxissymetric}). The 1/1 Lindblad resonance calculated this way is shown by a thick red curve on the negative half-plane in Fig.\,\ref{fig:VRtopologyANDmap}, left. The 1/1 Lindblad resonance on the dynamical map can be easily identified as a dominating black strip crossing the negative half-plane and intersecting the family of periodic orbits at  \mbox{$R \cong 4.6$\,kpc}.

\begin{figure}
\begin{center}
\epsfig{figure=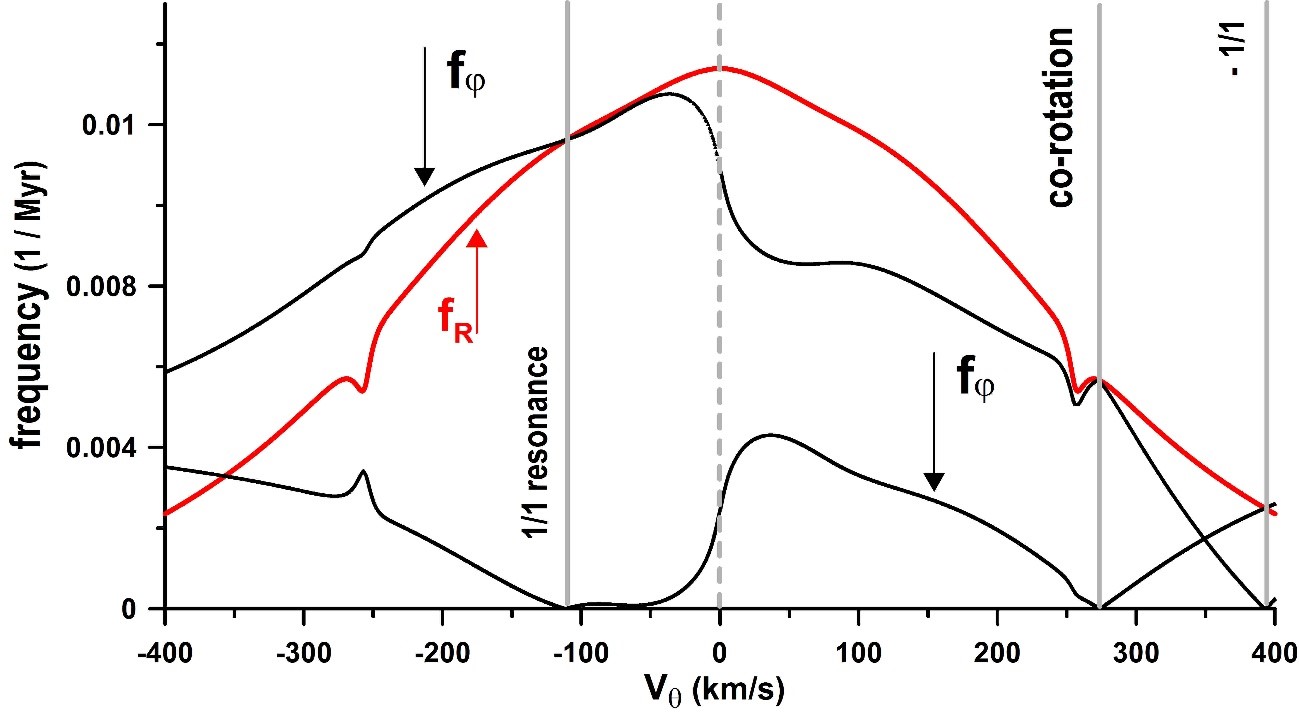,width=1.00\columnwidth ,angle=0}
\caption{Evolution of the unperturbed frequencies $f_R$ (red) and $f_\varphi$ (black) along $V_\theta$, calculated numerically with the initial values \mbox{$R=7.0$\,kpc}, \mbox{$p_R=0$\,km\,s$^{-1}$},  \mbox{$\varphi=90^\circ$},  and \mbox{$\zeta_0=0$} (see Eq.\,(\ref{eq:H1gaussian})) We note that $f_\varphi$  is presented by two branches. The location of the co-rotation and 1/1 resonances is shown by the vertical lines.
}
\label{fig:spec-negat}
\end{center}
\end{figure}

It is interesting to note in Fig.\,\ref{fig:VRtopologyANDmap}, right, that the 1/1 Lindblad resonance at \mbox{$V_\theta < 0$} seems to be a continuation of the co-rotation resonance at \mbox{$V_\theta > 0$}.  To understand the connection between the two resonances, we integrated numerically  the equations of motion of the full system\,(\ref{eq:eqsmotion}), setting the perturbation strength  $\zeta_0$ at zero. The dynamical power spectrum of this unperturbed problem was calculated along $V_\theta$ at the fixed value of the radius \mbox{$R=7$\,kpc} and is shown in Fig.\,\ref{fig:spec-negat}.

Two fundamental frequencies obtained are present in the spectrum: the radial frequency $f_R$ (red) and the azimuthal frequency $f_\varphi$ (black). The evolution of the radial frequency is continuous, similar for prograde and retrograde orbits, with local minima coinciding with the position of periodic orbits corresponding to the given value \mbox{$R = 7$\,kpc}.

The azimuthal frequency $f_\varphi$ is presented by two branches whose evolution seems to be independent. The existence of the two independent components in the variation of the angle $\varphi$ can be understood by analysing the corresponding equation of motion in Eqs.\,(\ref{eq:eqsmotion}), which shows the dependence on the two variables, $R$ and $J_\varphi$. The question which arises here is,  Which  of the two components of $f_\varphi$ is fundamental? For  \mbox{$V_\theta > 0$}, the correct answer  would be the lower branch, which tends to zero when the co-rotation resonance is crossed, at \mbox{$V_\theta = 273$\,km\,s$^{-1}$} (the vertical line labelled  `co-rotation' indicates this position). We note that, at the co-rotation resonance, the other component of $f_\varphi$ becomes equal to $f_R$ which, by definition, configures the 1/1 Lindblad resonance, as defined in this paper.

The dynamics is similar for the retrograde orbits, although the co-rotation motion is not expected for \mbox{$V_\theta < 0$} owing to the definition of the azimuthal angle as \mbox{$\varphi=\theta - \Omega_p\,t$}. In this case, the lower branch of  $f_\varphi$ tends to zero when the higher branch crosses the $f_R$--family at \mbox{$V_\theta = -109$\,km\,s$^{-1}$}. By definition, this event is associated with the 1/1 Lindblad resonance (the vertical line labelled  `1/1 resonance' indicates its position). Thus, the connection between the co-rotation and 1/1 resonances indicates the same dynamical nature of both, which is also supported by the same topology of the surfaces of section in Fig.\,\ref{fig:sos-all}.

Finally, there are other features of retrograde orbits which can be observed on the dynamical map in Fig.\,\ref{fig:VRtopologyANDmap}, right.  They are associated with Lindblad resonances of higher order, whose identification requires an additional study employing the models of resonant motion in action-angle variables.

\begin{figure*}
\begin{center}
\epsfig{figure=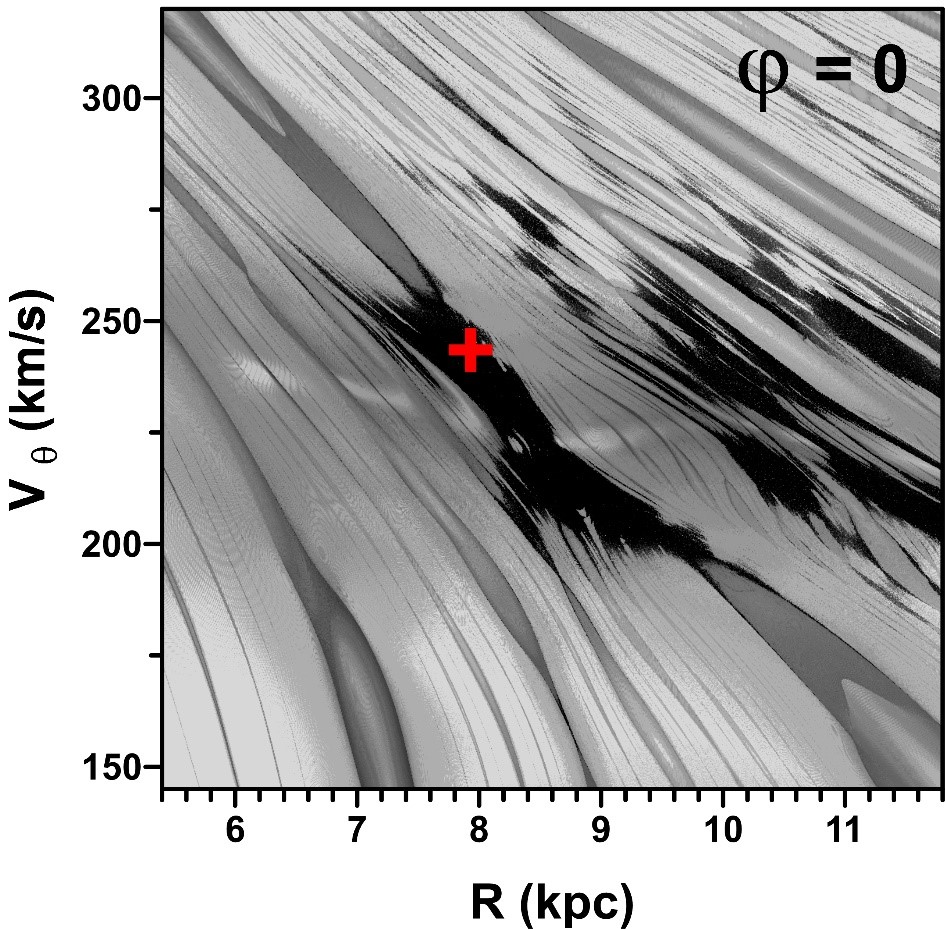,width=0.99\columnwidth ,angle=0}\quad
\epsfig{figure=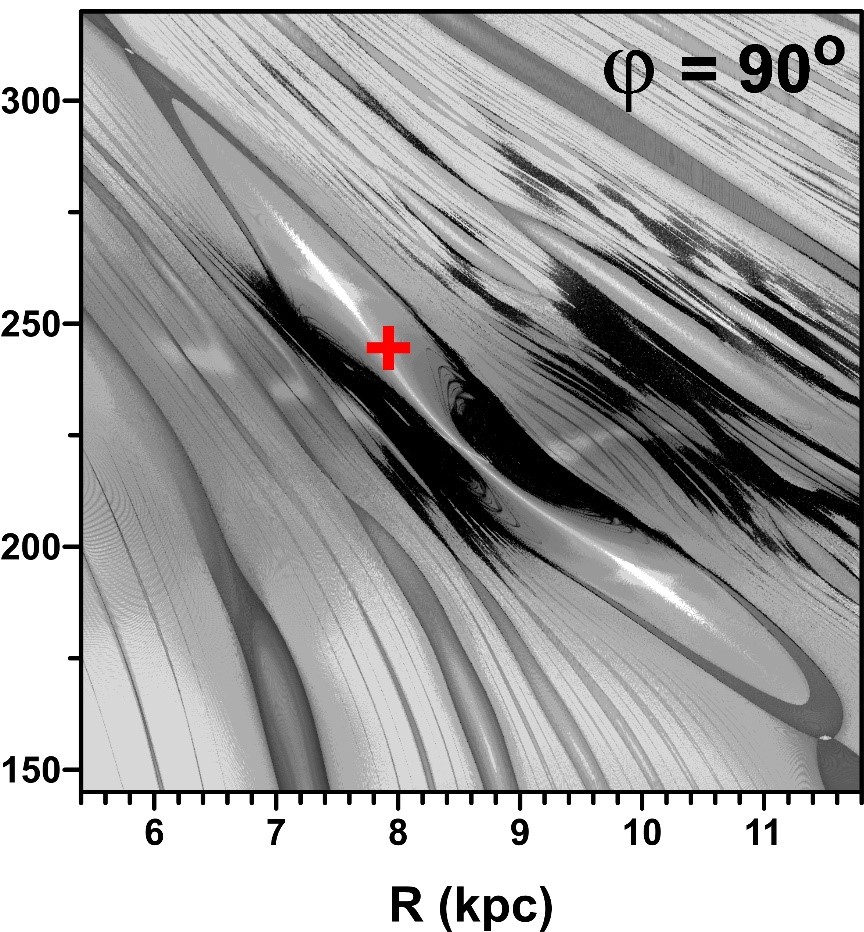,width=0.91\columnwidth ,angle=0}
\caption{
Left: Dynamical map of the neighbourhood of the co-rotation point calculated with the variables $p_R$ and $\varphi$ fixed at zero. The red cross shows the position of the SSP integrated backwards until \mbox{$p_R=0$}.
Right: Same as in the left graph, except \mbox{$\varphi=90^\circ$}. Comparing the graphs, we note that the star would be inside the co-rotation resonance island or in a very chaotic area, depending on the angle $\varphi$.
}
\label{fig:Dmap-corr}
\end{center}
\end{figure*}

\

\subsection{Dynamical portrait of the solar neighbourhood}
\label{sec:solar}

As shown in Fig.\,\ref{fig:sos-all}, top, the observations place the SSP close to the co-rotation point. To visualize the dynamical features of the domain surrounding this point, we construct two dynamical maps. In their construction, we use the 1000x1000 grid of initial conditions with \mbox{$p_R=0$}. We test two critical  values  of the co-rotation resonance: \mbox{$\varphi=90^\circ$} and \mbox{$\varphi=0$}. The map calculated with \mbox{$\varphi=0$} is shown in the left panel in Fig.\,\ref{fig:Dmap-corr}, while the map calculated with \mbox{$\varphi=90^\circ$} is in the right panel. The time series of $R$ was used in the construction of both maps.

We also plot the predicted position of the SSP on both planes in Fig.\,\ref{fig:Dmap-corr}. For this, the current initial conditions of the SSP  were integrated backwards in order to coincide with each plane at the condition \mbox{$p_R=0$}.

Comparing the two maps, we note that the current azimuthal position of the SSP has a significant influence on the qualitative behaviour  of its trajectory for a long time. In particular, if $\varphi$ is close to zero (left panel), the SSP will be in a region dominated by dynamical instabilities  which are associated with the separatrices of the  co-rotation resonance. This means that its behaviour for future times is unpredictable.

On the other hand, if $\varphi$ is close to $90^\circ$ (right panel), the SSP will evolve inside the stable region of the co-rotation resonance, as shown in the top panel in Fig.\,\ref{fig:sos-all}. Its motion on the $X$--$Y$ plane will be confined around the stable fixed point of the Hamiltonian (red point in Fig.\,\ref{fig:contoursBracos}). Assuming that the Sun is currently located in a void between two physical spiral arms, at  radius \mbox{$R_\odot = 8\,$kpc},  according to our model it is more likely  that the SSP would evolve on a stable orbit inside the co-rotation resonance. On the contrary, if we assumed its position close to \mbox{$\varphi=0$}, the SSP would be located very close to the arm, possibly lying inside of it, and evolving chaotically.

The angle $\varphi$ for the SSP can be estimated as follows. In the observed spiral pattern proposed by \citet{vallee2013IJAA}, the two arms nearest to the Sun are the Sagittarius-Carina arm, which passes about 1 kpc from the SSP at an inner  radius, and the Perseus arm,  at about 2 kpc from the SSP at an outer radius. Vall\'ee presents a four-arm spiral structure, while in our model we adopt  two arms. We  consider that one of the  arms in our model should coincide with either  the observed Sagittarius-Carina arm or with the Perseus arm. Since  both models use approximately the same pitch angle, we can match one of these two arms with one from our model by simple  rotation. We take as reference the point at which the co-rotation circle crosses the arm in the clockwise rotating Galaxy (with positive pitch angle, see Fig.~\ref{fig:contoursBracos}), and then calculate its angular position according to the above discussion. With this definition of $\varphi$, we find \mbox{$\varphi=$\,128$^\circ$} for the position of the Sun if we adopt the Sagittarius-Carina as one of the two arms in our model, and \mbox{$\varphi=$\,38$^\circ$} if the Perseus arm is adopted. However, since we use the negative value for the pitch angle in our calculations (see Table \ref{tab:1}), we must reflect these obtained $\varphi$--values with respect to the \mbox{$Y$-axis} in Fig.~\ref{fig:contoursBracos}, in order to obtain the correct values of $\varphi$ in our model. In this way, the azimuthal position of the SSP will be at \mbox{$\varphi=$\,52$^\circ$} or \mbox{$\varphi=$\,142$^\circ$} for the Sagittarius-Carina and Perseus arms, respectively. We believe that Sagittarius-Carina is the best choice since it is the most prominent and well-defined arm for most of the tracers.

\begin{figure}
\begin{center}
\epsfig{figure=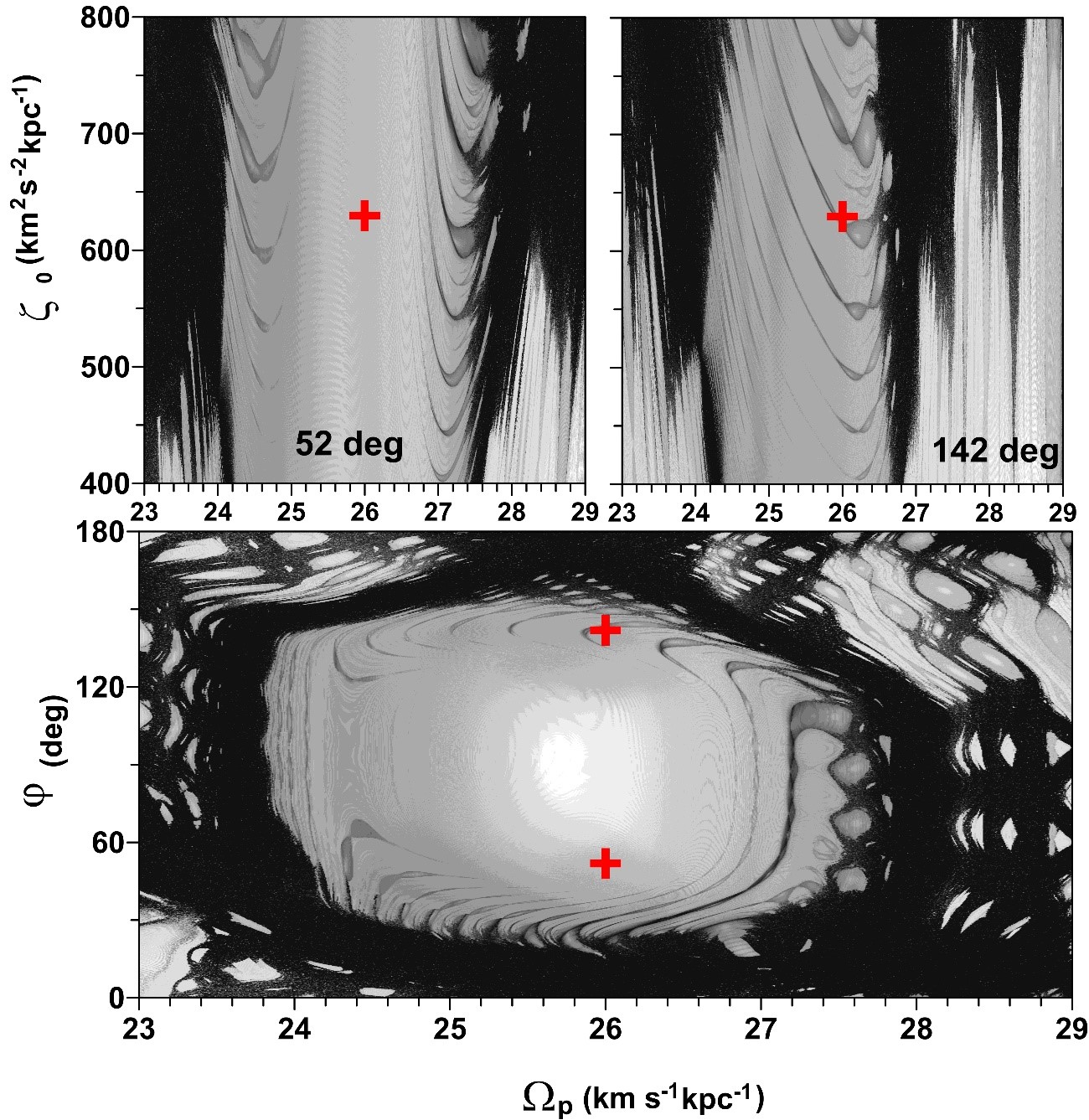,width=0.99\columnwidth ,angle=0}
\caption{Top: Dynamical maps on the parametric plane $\Omega_p$--$\zeta_0$,  defined by the pattern speed and the amplitude  of the spiral perturbation (Eq.~(\ref{eq:H1gaussian})), respectively. The maps were calculated for the initial conditions of the SSP and \mbox{$\varphi=52^\circ$} (left), \mbox{$\varphi=142^\circ$} (right), corresponding to the possible SSP azimuthal positions (see Sect.\,\ref{sec:solar}). The red crosses represent the location of the SSP according to the parameters used in this paper. The central light-coloured region corresponds to the stable co-rotation resonance.
Bottom: Same as in the top panel, except on the parametric plane $\Omega_p$--$\varphi$, defined by the pattern speed and the azimuthal angle, respectively.
}
\label{fig:MapParametersSSP}
\end{center}
\end{figure}

Figure\,\ref{fig:MapParametersSSP} shows three dynamical maps obtained for the initial conditions $R_0,\, p_{R,\odot},\,V_{\theta,\odot}$ corresponding to the SSP. The top panels show the plane $\Omega_p$--$\zeta_0$, calculated with the two different values of the azimuthal angle \mbox{$\varphi$}, $52^\circ$ and $142^\circ$. The red crosses represent the location of the SSP according to the parameters used in our study. The bottom panel shows the plane $\Omega_p$--$\varphi$, calculated with \mbox{$\zeta_0=630$}\,km$^2$\,s$^{-2}$\,kpc$^{-1}$. We draw two red crosses, corresponding to the two possible values of the SSP coordinate $\varphi$ described above. In all panels, the central light-coloured region  corresponds to the co-rotation resonance. The greyscale is the same as in Fig.\,\ref{fig:VRtopologyANDmap}. We see that the SSP remains inside the co-rotation resonance for values of $\Omega_p$ in the range \mbox{[24.5 -- 27]\,km\,s$^{-1}$\,kpc$^{-1}$}. For \mbox{$\varphi=52^\circ$}, this conclusion is valid for any value of $\zeta_0$ in the plotted range. For \mbox{$\varphi=142^\circ$}, however, larger values of $\zeta_0$ show secondary resonant islands associated with the co-rotation resonance. These results show that the SSP is still located inside the co-rotation resonance for small variations of the spiral amplitude $\zeta_0$, pattern speed $\Omega_p$, and azimuthal angle $\varphi$. For larger variations of these parameters, though, the SSP can either stay inside the co-rotation resonance or be in a chaotic zone. The possibility of being in a secondary resonance island associated with co-rotation is less likely.

\begin{figure}
\begin{center}
\epsfig{figure=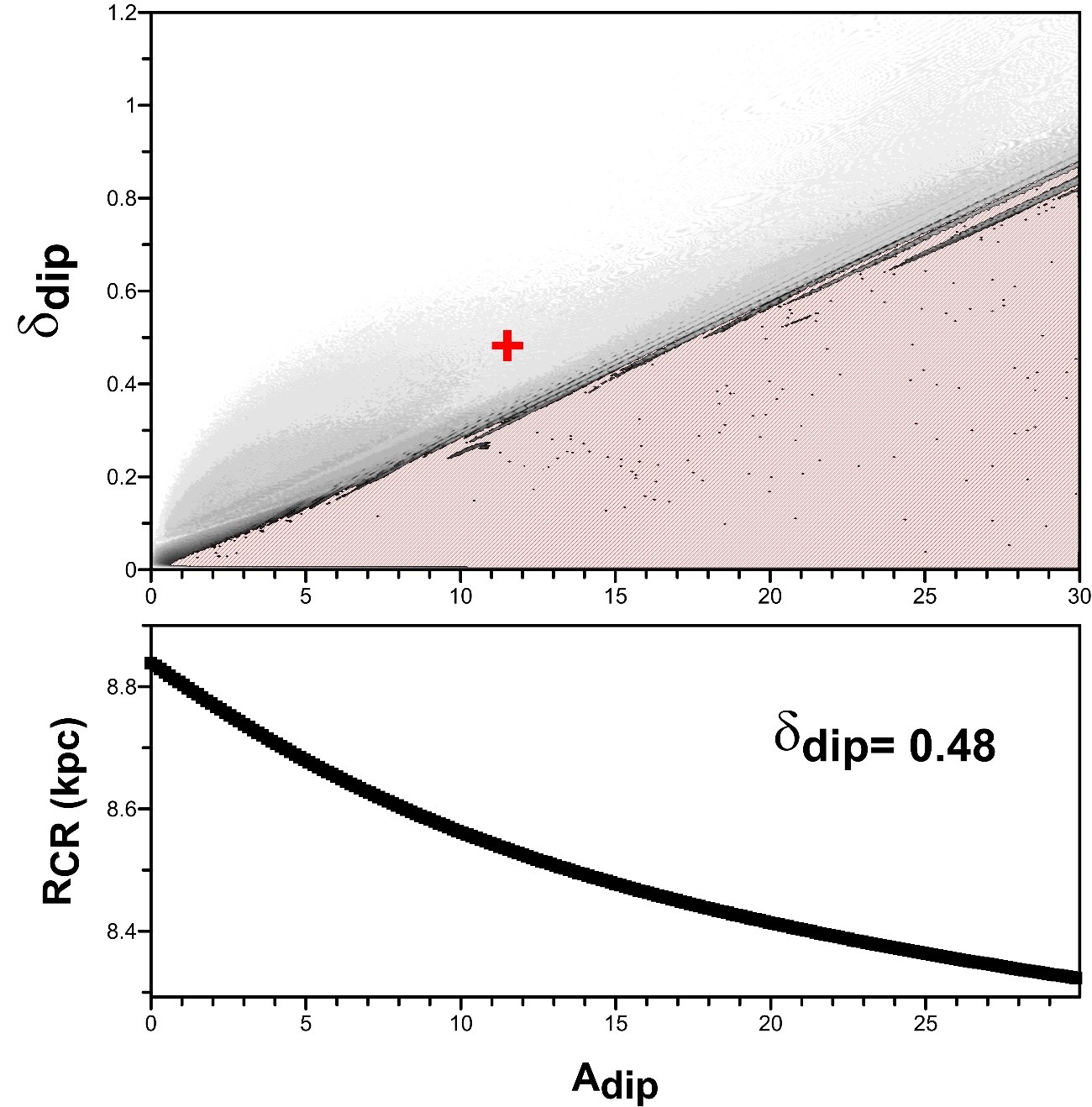,width=0.99\columnwidth ,angle=0}
\caption{Top: Parametric plane of the amplitude $A_{\textrm{dip}}$ (in km\,s$^{-1}$) and the thickness $\delta_{\textrm{dip}}$ (in kpc) of the co-rotation dip. The red cross indicates the values used in this paper. Bottom: Dependence of $R_{CR}$ on the dip's amplitude for a fixed value of $\delta_{\textrm{dip}}$. The red region corresponds to strong instabilities in the stellar motion.
}
\label{fig:MinimumParametric}
\end{center}
\end{figure}

\section{Co-rotation dip}\label{sec:corot_dip}

In this section we slightly modify the potential studied in the previous sections, varying the parameters of the last term  in Eq.\,(\ref{eq:Vrot}). Hereafter, we refer to this term as \emph{the co-rotation dip} and rewrite it in terms of new parameters, in the form
\begin{equation}\label{eq:corot-dip}
f_{\mathrm{dip}}(R) = -A_{\mathrm{dip}}\,\exp\left[-\frac{1}{2}\left(\frac{R-R_{\mathrm{dip}}}{\delta_{\mathrm{dip}}}\right)^2\right]\,,
\end{equation}
\noindent
where $A_{\mathrm{dip}}$ (in units of km s$^{-1}$), $\delta_{\mathrm{dip}}$, and $R_{\mathrm{dip}}$ (both in units of kpc) are the amplitude, the half-width, and the mean radius of the dip, respectively. The co-rotation dip corresponds to the small-amplitude valley which can be observed in the adopted form of the rotation curve in Fig.\,\ref{fig:Vrot}, near \mbox{$R=9$\,kpc}. In the literature, some authors  have also included a similar term related to the velocity dip to fit the observed rotation curve of the Galaxy traced by H\,{\scriptsize II} regions (e.g. \citealt{clemens1985ApJ, sofueHonmaOmodaka2009PASJ}). However, \citet{binneyDehnen1997MNRAS} showed that the form of the rotation curve for \mbox{$R>R_0$} as traced by H\,{\scriptsize II} regions is strongly affected by the uncertainties on the distances of these sources. The above-mentioned velocity dip could then be an artefact of this method of deriving the rotation curve for \mbox{$R>R_0$}.

In the present paper, we employ maser sources, with accurately determined distances and velocities, to trace the rotation curve beyond $R_0$ (see Sect.\,\ref{sec:rotcurve}). The maser sources provide  strong evidence for a velocity dip after $R_0$, which makes us believe  that this feature is real. Moreover, there is both theoretical and observational evidence for a deficit in the densities of stars and gas in the Galactic disk (in the co-rotation region). Such a deficit is intrinsically linked to the presence of the rotation velocity dip. For instance,
\citet{amoresLepineMishurov2009MNRAS} observed voids of H\,{\scriptsize I} gas density distributed in a ring-like structure with mean radius slightly outside the solar circle. \citet{zhang1996ApJ}, based on N-body simulations, showed that secular processes of energy and angular momentum transfer between stars and the spiral density wave induce the formation of a minimum in the stellar density, centred at the co-rotation radius.  Based on numerical integrations of test-particle orbits in a spiral potential, \citet*{lepineMishurovDedikov2001ApJ}, and also \citet{barrosLepineJunqueira2013MNRAS} observed the formation of minima of density around the co-rotation radius of the simulated disks. \citet{sofueHonmaOmodaka2009PASJ} employed a ring density structure in their Galactic disk model to fit the observed dip in the rotation curve at \mbox{$R\sim 9$} kpc.

This dip near co-rotation influences the stability of the orbits inside the co-rotation region. To show this, we test different values for dip amplitudes and widths in order to construct the parametric plane shown in Fig.\,\ref{fig:MinimumParametric}. For each pair (A$_{\textrm{dip}}$, $\delta_{\textrm{dip}}$), the co-rotation radius was calculated looking for equilibrium solutions of the full Hamiltonian\,(\ref{eq1}) described in Sect.\,\ref{sec:spiralbranches}. The corresponding orbit was integrated numerically and SAM-analysed.  The spectral number obtained was plotted on the parametric plane using a greyscale code. Light grey regions correspond to stable co-rotation solutions, whereas black regions correspond to chaotic co-rotation resonances. The red domain in Fig.\,\ref{fig:MinimumParametric} corresponds to large instabilities in the stellar motion. This region coincides with the region where \mbox{$\kappa^2(R_{CR}) < 0$}  (see Eq.\,(\ref{eq3-1})), which means hyperbolic equilibrium solutions of ${\mathcal H}_0$, when the amplitude of the spiral perturbation tends to zero.

Our choice of the parameters \mbox{A$_{\textrm{dip}}= 11.86$\,km\,s$^{-1}$} and \mbox{$\delta_{\textrm{dip}}= 0.48$\,kpc}, marked by a red cross on the parametric plane in Fig.\,\ref{fig:MinimumParametric}, places the co-rotation solution inside the stable region. In the case when the amplitude of the velocity dip beyond the solar radius \mbox{$R_0=8$\,kpc} is larger or its width is smaller, the co-rotation region becomes chaotic. It is worth noting, however, that the values of the parameters were determined through the fitting of the observed rotation curve of the Galaxy and using the specific exponential function\,(\ref{eq:corot-dip}) to describe the co-rotation dip. Thus, the region around co-rotation can be considered  stable in the framework of our model and for the adopted set of the parameters.

\section{Conclusions}

In this paper, we presented a new approach to the study of the dynamics of stars in spiral galaxies, based on widely used methods of celestial mechanics \citep{michtchenkoEtal2002Icar, ferrazmeloMichtchenkoEtal2005LNP}.

First, we modeled the rotation curve of our Galaxy based on observations of H\,{\scriptsize I}, CO, and maser sources. The rotation-curve model provides the axisymmetric component of the Galactic potential in the mid-plane of the disk. In this way, the potential model relies directly on rotation curve data, without any assumption of a mass-model description of the Galaxy's components.

Our model is limited by the extent to which the fit to the rotation curve is reliable and by the existence of massive bodies orbiting the Galaxy, and thus will be valid up to a radius of \mbox{$R\sim 30\,$kpc}. The inner region (\mbox{$R<3\,$kpc}) may also be excluded from the physical point of view, since there is evidence of the presence of a bar in this central region.

We employed a spiral potential described by Gaussian-shaped potential wells \citep{junqueiraEtal2013AA}, and re-scaled its parameters in order to be consistent with the parameters of the LSR, adopted in this paper, \mbox{$R_0=8\,$kpc} and \mbox{$V_0=230\,$km\,s$^{-1}$}. We believe that the Gaussian-grooved model is a more accurate description of the azimuthal profile of the spiral arms than the classical cosine perturbation. Indeed, as shown in \citet{junqueiraEtal2013AA}, a number of observed structures in the Milky Way can be associated with resonant orbits predicted by such a Gaussian spiral potential model.

We then introduced the representative plane \mbox{$R$--$V_\theta$} of initial configurations, with fixed \mbox{$p_R=0$} and \mbox{$\varphi=90^\circ$}, to explore the stellar dynamics described by the adopted spiral-galaxy model. This plane is representative of the whole phase space of the system and its analysis provides the precise positions of the co-rotation and Lindblad resonances, as well as the degree of chaoticity of stellar orbits. It is also suitable for the analysis of existent and upcoming data regarding the peaks in the distribution function of stars in the Galactic phase space, such as kinematic moving groups \citep{antojaEtal2008AA, antojaEtal2010LNEA, smith2016ASSL}.

The analysis of the dynamics was done by means of \emph{dynamical maps} based on the SAM approach \citep{michtchenkoEtal2002Icar, ferrazmeloMichtchenkoEtal2005LNP}. A complementary study was done by means of \emph{dynamical power spectra} of the orbits corresponding to curves $V_\theta(R)$ on the representative plane, particularly along spiral branches and along the rotation curve. This method allows us to identify the fundamental frequencies of the problem. The comparison with theoretical results in the axisymmetric potential permits us to correctly classify resonances in phase space. The results obtained agree with the predictions of classical Lindblad resonances given by the epicyclic approximation.  The concept of Lindblad resonances was then extended over the whole phase space of the system.

We also analysed the solar neighbourhood on the representative plane of initial conditions, showing that a star with solar kinematic parameters is likely to evolve inside the co-rotation resonance. Depending on its current value of $\varphi$, it may be either in a stable or in a chaotic zone. The analysis of the co-rotation dip, present in the adopted model of the rotation curve, shows that our choice of parameters for the rotation curve corresponds to a stable co-rotation resonance, whereas slightly different parameters could drastically change the dynamics around co-rotation.

Finally, the approach presented in this work describes with high precision the resonance chains in the whole phase space and provides a more complete dynamical portrait than the classical epicyclic approximation. Although many stars in the solar neighbourhood seem to be on quasi-circular orbits, there is a great deal of  evidence of non-circular motion; there is also evidence of  retrograde motion  \citep{carneyASPCS1993}, for which our model is more suitable. Thus, our approach may have an impact on the identification of dynamical signatures of  resonant orbits and on the degree of chaos in the solar neighbourhood, and on its local phase-space structure \citep{chakrabarty2007AA,chakrabartySideris2008AA}. It may also shed light on the origin of kinematic moving groups, believed to be related to resonances \citep{antojaEtal2008AA,antojaEtal2009ApJL,antojaEtal2011MNRAS,
morenoPichardoSchuster2015MNRAS,martinezmedinaEtal2016ApJL}. These topics, as well as a dynamical analysis of three-dimensional potentials, will be the subject of a forthcoming study.


\begin{acknowledgements}

This work was supported by the São Paulo State Science Foundation, FAPESP, and the Brazilian National Research Council, CNPq. RSSV acknowledges the financial support from FAPESP grant 2015/10577-9. DAB acknowledges support from the Brazilian research agency CAPES, under the program PNPD. This work has made use of the facilities of the Laboratory of Astroinformatics
(IAG/USP, NAT/Unicsul), whose purchase was made possible by FAPESP (grant 2009/54006-4) and the INCT-A.
We acknowledge the anonymous referee for the detailed review and for the many helpful suggestions which allowed us to improve the manuscript.

\end{acknowledgements}




\appendix

\section{Comparison with the  cosine profile of the spiral arms}\label{app1}

Many works treat spiral arms as perturbations with the azimuthal dependence in the cosine form, first introduced by \citet{linShu1964ApJ} and then largely adopted in the literature \citep[e.g.][and references therein]{contopoulos1973ApJ, papayannopoulos1979aAA, papayannopoulos1979bAA, chakrabartySideris2008AA, danielWyse2015MNRAS}.

\begin{figure}
\begin{center}
\epsfig{figure=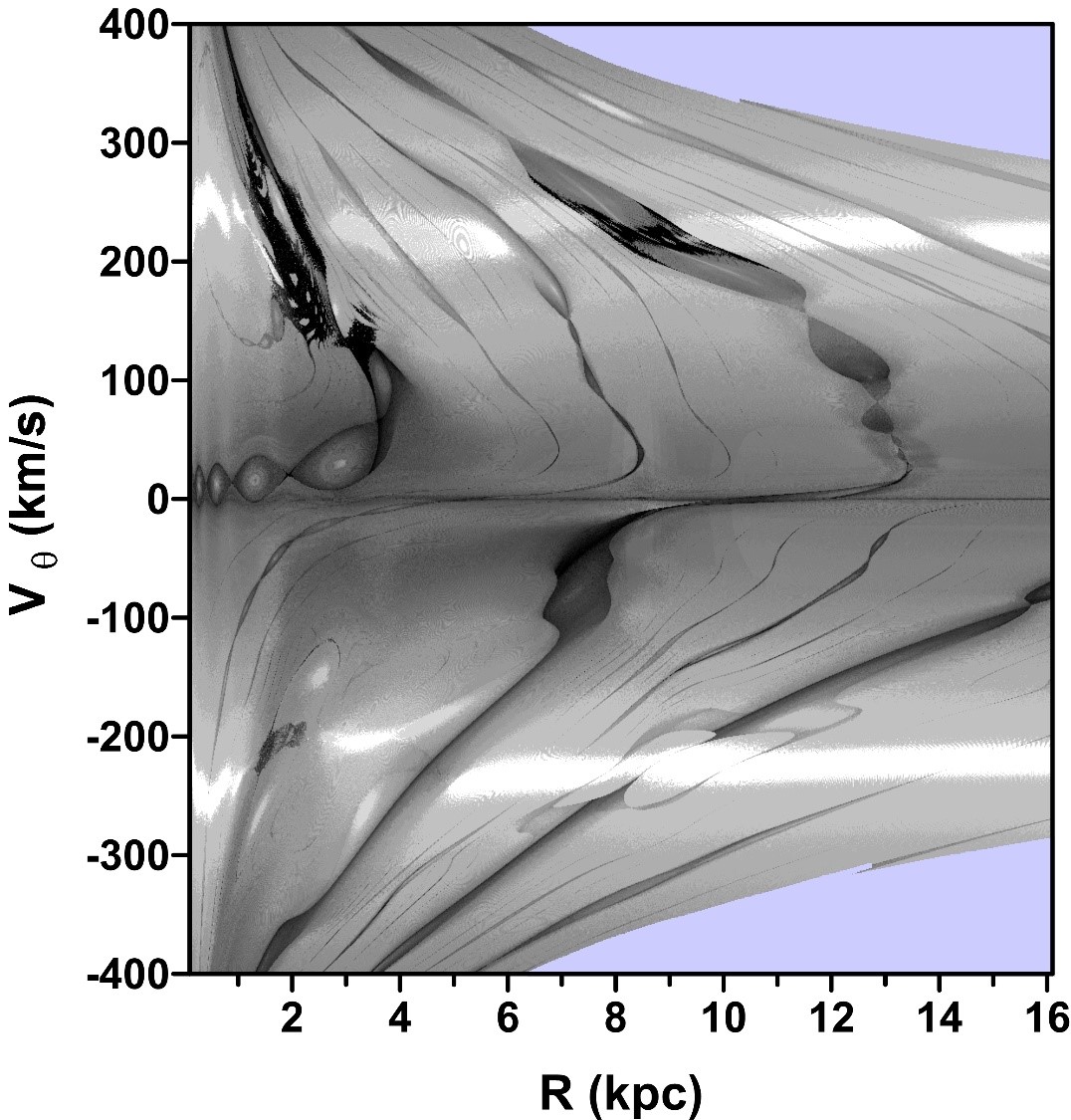,width=0.99\columnwidth ,angle=0}
\caption{Dynamical map of the representative plane of initial conditions, for the cosine perturbation given in Eq.\,(\ref{eq:PhiCosine}) calculated for the time series $J_\varphi(t)$ and the fixed angle \mbox{$\varphi=90^\circ$}. The blue shaded regions correspond to the initial conditions of the orbits which go beyond 30 kpc, regions where we consider that our model is no longer applicable.
}
\label{fig:VRmapCosine}
\end{center}
\end{figure}

Figure \ref{fig:VRmapCosine} shows the $R$--$V_\theta$ dynamical map calculated with the perturbation $\Phi_{1,\rm cos}(R,\varphi)$ given in  Eq.\,(\ref{eq:PhiCosine}), for the same parameters listed in  Table\,\ref{tab:1}. We see that the resonance chains remain at approximately the same radii (as expected from the non-perturbed calculations) as those shown in Fig.\,\ref{fig:VRtopologyANDmap}; however, the shape and degree of chaos of the resonances is different: while the Gaussian-grooved spiral perturbation\,(\ref{eq:H1gaussian}) gives strips that are more highly defined for the resonances, the cosine perturbation shows high chaoticity for small-radii resonances. Outside co-rotation, however, the groove-shaped perturbations give rise to more chaotic orbits.

Particular attention must be paid to the inner 2/1 Lindblad resonance (ILR). While the ILR for the cosine perturbation shows strong chaotic behaviour, as we see in Fig.\,\ref{fig:VRmapCosine}, the groove-shaped spirals seem to give rise to robust resonance islands, as we see for the ILR in Fig.\,\ref{fig:VRtopologyANDmap}. We can conclude that cosine perturbations tend to generate a wider chaotic region near the ILR,
especially close to the rotation curve. The width of the ILR is also greater than in the Gaussian-shaped spiral potential, as we can see by comparing Figs.\,\ref{fig:VRtopologyANDmap} and \ref{fig:VRmapCosine}.

The other resonances of the system are more pronounced in the Gaussian-grooved case. These results may be related to the differences in the azimuthal profiles of the two spiral potentials. But they may also be related to the differences in the radial dependence of the width of the spirals in the two models: the Gaussian-shaped spiral arms are modelled with a constant width along the Galactic disk, whereas in the cosine perturbation it increases with radius.

\section{Analytical and numerical techniques}\label{app2}

\subsection{Motion in the unperturbed system: Azimuthal and radial frequencies}\label{sec:frequenciesUnperturbed}

For completeness, in this  appendix we present the equations used in the analytical predictions of resonances. We also compare the analytical predictions with the predictions from the epicyclic approximation, showing how quantities calculated from orbits which are far from circular are approximated by the corresponding quantities predicted for nearly circular orbits. The epicyclic approximation is not valid if deviations from circular motion become non-negligible, as we see for instance in Figs.\,\ref{fig:VRtopologyANDmap} and \ref{fig:specJ1Omegap}.

The orbital ($\theta$-) and radial ($R$-) periods of a regular bounded orbit can be estimated using the axially symmetric potential $\Phi_0$
\citep{binneytremaineGD}.
The angular momentum $J_\varphi$ and the energy
\begin{equation}\label{eq:Eunperturbed}
E=\frac{1}{2}\dot R^2 + \Phi_0(R)+\frac{J_\varphi^2}{2R^2}
\end{equation}
are integrals of motion which can be calculated from the orbit's initial conditions. It can be shown that
\citep[eq. 3.18a]{binneytremaineGD}
\begin{equation}\label{eq:TRaxissymetric}
 T_R= 2\int_{R_1}^{R_2} \frac{1}{\dot R}\, dR
 = 2\int_{R_1}^{R_2} \frac{dR}{\sqrt{2[E-\Phi_0(R)]-(J_\varphi)^2/R^2}},
\end{equation}
where the turning points $R_1$ and $R_2$ are the solutions of Eq.\,(\ref{eq:Eunperturbed}) with \mbox{$\dot R=0$}.
We  also have that the azimuthal period $T_\theta$, in the inertial frame, is
\citep[eq. 3.19]{binneytremaineGD}
        \begin{equation}
         T_\theta=\frac{2\pi}{|\Delta\theta|} T_R,
        \end{equation}
where the azimuthal angle variation $\Delta\theta$ during this radial oscillation is
\citep[Eq. 3.18b]{binneytremaineGD}
\begin{equation}\label{eq:DeltaThetaaxissymetric}
 \Delta \theta = \oint \dot\theta\, dt
 = 2J_\varphi\int_{R_1}^{R_2}\frac{dR}{R^2\sqrt{2[E-\Phi_0(R)]-(J_\varphi)^2/R^2}}.
\end{equation}
In the rotating frame, \mbox{$\Delta\varphi=\Delta\theta-\Omega_p T_R$}. We then have
\begin{equation}
T_R=\frac{|\Delta\varphi|}{2\pi}T_\varphi.
\end{equation}
Resonances in this approximation then occur for
\begin{equation}\label{eq:resonanceEstimative}
 \frac{\Delta\varphi}{2\pi}=\frac{j}{n} \quad\Rightarrow\quad
 \Delta\theta-\Omega_p T_R=2\pi\,\frac{j}{n},
\end{equation}
with $n,j$ coprime integers and $\Delta\theta$ and $T_R$ given by the equations above.

It is usual to approximate the above quantities by their linearized values along the rotation curve, that is:
\mbox{$\Omega_\theta\approx\Omega$} (i.e., \mbox{$\Omega_\varphi\approx\Omega-\Omega_p$}) and \mbox{$\Omega_R\approx\kappa$}, where the angular velocity of circular orbits $\Omega(R)$ is given by \mbox{$\Omega(R) = V_{\rm rot}/R$}
and $\kappa(R)$ is the (radial) epicyclic frequency, defined by the equation \mbox{$\dot{p}_R=-\kappa^2(R_{\rm c})\cdot (R-R_{\rm c})$} for each $R_{\rm c}$ \citep{binneytremaineGD}. It is then given by (\citealp{binneytremaineGD}, eq. 3.79a)
\begin{equation}\label{eq3-1}
\kappa^2(R_{\rm c})=
\frac{\partial^2\Phi_0}{\partial R^2} + \frac{3}{R}\frac{\partial\Phi_0}{\partial R}\Bigg|_{R_{\rm c}}.
\end{equation}

The ``epicyclic approximation'' is therefore defined by the linearized problem:
\mbox{$\Omega_\varphi=\Omega-\Omega_p$} and \mbox{$\Omega_R=\kappa$}. The resonances which appear in this approximation are known as \emph{Lindblad resonances}. They are given by
\begin{equation}\label{eq:LindbladRess}
\Omega-\Omega_p=\frac{j}{n}\kappa,
\end{equation}
with $j$ and $n$ coprime integers.
Assuming \mbox{$\Delta\varphi>0$} (see Eq.\,(\ref{eq:resonanceEstimative})), we have
\begin{equation}\label{eq:frequenciesComparison}
 \frac{\Omega_\varphi}{\Omega_R}=\frac{T_R}{T_\varphi}
 = \frac{\Delta\theta-\Omega_p T_R}{2\pi}
 = \bigg(\frac{T_R}{2\pi}\bigg)\cdot\bigg[\frac{\Delta\theta}{T_R}-\Omega_p\bigg]
 \approx \frac{1}{\kappa}\cdot[\Omega-\Omega_p],
\end{equation}
where in the last term we approximate the frequencies by the corresponding nearly circular (epicyclic) motion. It is worth remarking that this approximation is only valid for orbits with small deviations from the rotation curve. Therefore, the resonance radii given by the epicyclic approximation (\ref{eq:LindbladRess}) need not  be equal to the resonance radii obtained from the full numerical, perturbed calculations, or  to the axisymmetric estimations (\ref{eq:resonanceEstimative}) if the corresponding orbit is not nearly circular.  On the other hand, the unperturbed approximation (see Eqs.\,\ref{eq:TRaxissymetric}--\ref{eq:resonanceEstimative}) describes the positions of resonances with good accuracy, as shown in the text.

\subsection{Numerical techniques: Spectral Analysis Method}\label{sec:SAM}

The stellar motion in the full galactic potential (given by Eq.\,(\ref{eq1})) is analysed by purely numerical techniques applying the Spectral Analysis Method  (\citealp{michtchenkoEtal2002Icar, ferrazmeloMichtchenkoEtal2005LNP}), frequently used in planetary and asteroidal sciences.

The SAM is used to distinguish between regular and chaotic domains on the phase space of Hamiltonian systems and is based on the well-known features of power spectra \citep{powell1979}. It involves two main steps. The first step is the numerical integration of the equations of motion defined by the Hamiltonian\,(\ref{eq1}), while the second step consists of the spectral analysis of the output of the numerical integrations. The series giving the time variation of stellar orbital elements (e.g. the canonical phase-space coordinates) are Fourier-transformed using a standard fast Fourier transform (FFT) algorithm.

The Fourier transforms of the output make it possible to distinguish between regular and irregular motions. Indeed, regular motions are conditionally periodic and  any orbital element $ele(t)$ depends on time as a function
\begin{equation}\label{fft}
  ele(t) = \sum_\textbf{k}{A_\textbf{k} \exp({2\pi i\,\textbf{k}\cdot\textbf{f}\, t}}),
\end{equation}
where \textbf{f} is a frequency vector whose components are the fundamental frequencies of motion and \mbox{$\textbf{k} \in\textbf{Z}^D$}, where $D$ is the dimension of the configuration space (or the number of degrees of freedom) of the problem. When the independent frequencies are constant in time, the power spectrum of $ele$, obtained from its Fourier transform, consists of the lines associated with the independent frequencies, whose number is equal to the number of degrees of freedom of the dynamical system, and with those corresponding to higher harmonics and linear combinations of the independent frequencies. The number of peaks that are above an arbitrarily defined “noise” level is defined as the spectral number $N$ and is associated with this $ele$. In this paper, we define the noise level as 5\% of the largest peak in the spectrum.

Chaotic motions are no longer conditionally periodic and the fundamental frequencies of the system vary in time.  The power spectrum of chaotic motion is not discrete, showing broadband components; therefore, the spectral number $N$ will be characterized by very high values in this case.

Thus, the spectral number $N$ can be used to quantify the chaoticity of the system in the following way: small values of $N$ correspond to regular motion, while large values of $N$ indicate the onset of chaos. Using this criterion, we construct \emph{dynamical maps} of the system under study.

Dynamical maps associate a spectral number $N$ of a given orbital element to each point on the plane of initial conditions by means of the spectral analysis of the corresponding orbit described above. They are useful tools in the characterization of the phase space and identification of the principal dynamical mechanisms acting on stellar motion.  Once $N$ is determined for all initial conditions on a grid covering the plane of initial conditions, we plot it using a greyscale that varies logarithmically from white (\mbox{$N=1$}) to black ($N$ maximum). Since high values of $N$ indicate the onset of chaos, the grey tones indicate different degrees of stochasticity of solutions with initial conditions starting on the plane: lighter regions correspond to regular motion and darker tones indicate chaotic motion.

The spectral analysis allows an efficient identification of the main oscillations  contained  in  the  trajectory of the star. Power spectra are plots of the amplitude of the Fourier transform against frequency. In order to see how the spectra change when initial conditions vary, we adopt a \emph{dynamic power spectrum}, which is constructed  plotting the frequencies of the peaks on the power spectra as functions of one parameter describing a family of solutions.  Over the domains of regular motion, the proper frequencies vary continuously when the initial conditions are gradually varied. When resonances are approached, the topology of the phase space suffers qualitative transformations and the frequency evolution shows a discontinuity characterized by the erratic scatter of values when chaotic layers associated with separatrices are crossed. In systems with two degrees of freedom, where the chaoticity may be studied with the help of surfaces of section (Poincar\'e maps), dynamic power spectra allow us to understand the dynamics in areas where the maps show intricate features or where the features are too thin to be visible.

It is worth emphasizing that the SAM is simple to implement. Its performance, tested on the standard map problem with known solutions, is robust (for details, see Ferraz-Mello et al. 2005, Appendix A). Based on a solid criterion, the method is infallible in distinguishing regular from chaotic motion. The resolution of the SAM is defined by the chosen grid: for smaller spacings between the grid nodes, the structure of the dynamical map exhibits finer details. Also, since the independent modes of motion of the problem under study may show distinct behaviour, it is advisable to calculate and visualize the spectral number $N$ for all variables of the problem.

The spectral number $N$ also depends  on the integration time span; in other words, an orbit classified as regular can appear as chaotic if a larger time span is used in the integrations. Indeed, if the diffusion rate of the main frequencies is below the Fourier  transform resolution (defined by the time span), the spectral analysis methods are unable to detect chaos. Thus, the chosen total integration time should be  large enough to  allow the  chaos generated by Lindblad resonances to be distinguished. On the other hand, the dependence of $N$ on the noise-level is not critical, and the choice of 5\% of the largest peak in the spectrum  yields good results.

\end{document}